# Controlling the domain structure of ferroelectric nanoparticles using tunable shells

Anna N. Morozovska [1*], Eugene A. Eliseev[2], Yevhen M. Fomichov[3], Yulian M. Vysochanskii[4], Victor Yu. Reshetnyak[5†], and Dean R. Evans[6‡]

[1] Institute of Physics, National Academy of Sciences of Ukraine, 46, pr. Nauky, 03028 Kyiv, Ukraine

[2] Institute for Problems of Materials Science, National Academy of Sciences of Ukraine, Krjijanovskogo 3, 03142 Kyiv, Ukraine

[3] Charles University in Prague, Faculty of Mathematics and Physics, V Holešovičkach 2, Prague 8, 180 00, Czech Republic

[4] Institute of Solid State Physics and Chemistry, Uzhgorod University, 88000 Uzhgorod, Ukraine

[5] Taras Shevchenko National University of Kyiv, Volodymyrska street 64, Kyiv, 01601, Ukraine

[6] Air Force Research Laboratory, Materials and Manufacturing Directorate, Wright-Patterson Air Force Base, Ohio, 45440, USA

## Abstract

The possibility of controlling the domain structure in spherical nanoparticles of uniaxial and multiaxial ferroelectrics using a shell with tunable dielectric properties is studied in the framework of Landau-Ginzburg-Devonshire theory. Finite element modeling and analytical calculations are performed for $Sn_2P_2S_6$ and $BaTiO_3$ nanoparticles covered with high-k polymer, temperature dependent isotropic paraelectric strontium titanate, or anisotropic liquid crystal shells with a strongly temperature dependent dielectric permittivity tensor. It appeared that the "tunable" paraelectric shell with a temperature dependent high dielectric permittivity (~300 – 3000) provides much more efficient screening of the nanoparticle polarization than the polymer shell with a much smaller (~10) temperature-independent permittivity. The tunable dielectric anisotropy of the liquid crystal shell (~ 1 – 100) adds a new level of functionality for the control of ferroelectric domains morphology (including a single-domain state, domain stripes and cylinders, meandering and labyrinthine domains, and polarization flux-closure domains and vortexes) in comparison with isotropic paraelectric and polymer shells. The obtained results indicate the opportunities to control the domain structure morphology of ferroelectric nanoparticles covered with tunable shells, which can lead to the generation of new ferroelectric memory and advanced cryptographic materials.

* Corresponding author 1: anna.n.morozovska@gmail.com

† Corresponding author 2: victor.reshetnyak@gmail.com

‡ Corresponding author 3: dean.evans92@gmail.com

## I. INTRODUCTION

Nanosized ferroelectrics attract permanent attention of researchers as unique model systems for fundamental studies of polar surface properties, various screening mechanisms of spontaneous polarization by free carriers, and possible emergence of versatile multi-domain states [1, 2, 3, 4, 5, 6]. One of the most promising objects are ferroelectric nanoparticles, which are relatively easy to synthetize and control their polar properties. Classical examples are nontrivial experimental results of Yadlovker and Berger [7, 8, 9], who revealed and studied the ferroelectricity enhancement in Rochelle salt cylindrical nanoparticles. Frey and Payne [10], Zhao et al. [11], Zhu et al. [12], Drobnich et al. [13], Erdem et al. [14], Shen et al. [15], and Golovina et al. [16, 17, 18] demonstrated the possibility to control the phase transition temperatures and other peculiarities of $BaTiO_3$, $Sn_2P_2S_6$, $PbTiO_3$, $SrBi_2Ta_2O_9$**,** and $KTa_{1-x}Nb_xO_3$ nanopowders and nanoceramics by finite size effects.

The continuum phenomenological Landau-Ginzburg-Devonshire (**LGD**) approach combined with the electrostatic equations allows one to establish the physical origin of the anomalies in the polar and dielectric properties of ferroelectric nanoparticles, and predict the changes of their phase diagrams with size reduction. For instance, using the LGD approach Niepce [19], Huang et al. [20, 21], Lin et al. [22], Glinchuk et al. [23, 24, 25], Ma [26], Khist et al. [27], Wang et al. [28, 29, 30], Morozovska et al. [31, 32, 33, 34], and Eliseev et al. [35, 36, 37] have shown that the transition temperatures, the degree of spontaneous polarization or/and magnetic ordering in spherical, ellipsoidal, and cylindrical nanoparticles of sizes (4 – 100) nm are conditioned by finite size effects. The various physical mechanisms ruling the size effects are surface tension, correlation effect, depolarization field originated from the incomplete screening of spontaneous polarization, flexoelectricity, electrostriction, magnetoelectric coupling, magnetostriction, rotostriction, and Vegard-type chemical pressure.

Analytical description of domain structure morphology and phase diagrams of uniaxial ferroelectric nanoparticles was proposed in the framework of LGD approach by Eliseev et al. [38] and Morozovska et al. [39]. To model realistic conditions of incomplete screening of spontaneous polarization at the particle surface, it was considered that the particle is covered with an ultra-thin layer of screening charge with an effective screening length. The phase diagrams, calculated by finite element modelling (**FEM**), demonstrate the emergence of poly-domain region at the tricritical point and its broadening with increasing the screening length for the particle radius above a critical value.

Metastable and stable labyrinthine domain structures can be formed in $Sn_2P_2S_6$ and $CuInP_2S_6$ nanoparticles with a radius (8-10) nm and greater.

Gregg [40] reviewed experimental and theoretical works to find out whether or not complex arrangements of dipoles (such as flux-closure, vortex, or skyrmion patterns) exist in multiaxial ferroelectrics. Metastable meandering domain walls can exist in thin $BiFeO_3$ films [41, 42]. Rodriguez et al. [43] experimentally studied two-dimensional arrays of ferroelectric lead zirconate titanate nanodots with the help of piezoresponse force microscopy. Obtained results allowed them to suggest the presence of quasi-toroidal polarization ordering. Karpov et al. [44] used Bragg coherent diffractive imaging of a single $BaTiO_3$ nanoparticle in a composite polymer/ferroelectric capacitor to study the behavior of a three-dimensional vortex formed due to competing interactions involving ferroelectric domains. The investigation of structural phase transitions under the influence of an external electric field shows a mobile vortex core exhibiting a reversible hysteretic transformation path.

Mangeri et al. [45] simulated the behavior of polarization in isolated spherical $PbTiO_3$ and $BaTiO_3$ nanoparticles embedded in a dielectric medium. It was shown that the equilibrium polarization topology is strongly affected by particle diameter, as well as by the choice of inclusion and matrix materials, with single-domain, vortex-like and poly-domain patterns emerging for various combinations of size and materials parameters. Wang et al. [46] calculated the electrocaloric effect of $PbTiO_3$ nanoparticles with complicated vortex-like domain structure. Pitike et al. [47] used LGD phenomenology to model ferroelectric nanoparticles and have found that the critical particle sizes of the texture instabilities are strongly dependent on the particle shape. Zhu et al. [48] modeled polar properties of ferroelectric nanoparticles with different shapes and sizes. It was shown that large enough particles do not have a single-domain ground state supporting polarization field gradients at zero field; hence the process of polarization reversal occurs via the emergence of intermediate phases that involves an appreciable amount of vorticity.

To the best of our knowledge existing theoretical papers (cited above and many others) do not consider ferroelectric nanoparticles in a media with temperature-dependent dielectric and/or elastic properties. However, if the dielectric and/or elastic properties of the shell are temperature-dependent, the temperature variation results in the change of electric field inside and outside the core, which should influence on the ferroelectric polarization of the nanoparticle. It should be noted that there are several possible ways to change the value and anisotropy of the shell dielectric permittivity, e.g. to use thermo-responsive polymer gel [49], liquid crystal [50, 51], or elastomer [52] as a "**tunable shell**". Temperature-dependent lattice mismatch in core-shell nanoparticles occurs when the core and shell materials have different lattice parameters [53, 54, 55]. If there is a significant lattice mismatch at the core-shell interface, it results in an additional strain energy, which can affect the morphology of the domain structure and ferroelectric polarization in the core.

Despite that it is relatively easy to change the shell dielectric and/or elastic properties, the temperature evolution of polarization state and domain morphology in a ferroelectric nanoparticle ("core") covered with a tunable shell was not studied at all. It should be noted that the study can open new possibilities to control the domain structure morphology in ferroelectric nanoparticles covered with tunable shells, and can be useful for the next generation of ferroelectric memory and advanced cryptographic materials.

Motivated to fill the gap in knowledge, this work studies the temperature evolution of unusual domain states and phase diagrams in uniaxial and multiaxial ferroelectric nanoparticles ("core") covered with tunable semiconducting shells. Free carriers in the shell provide incomplete screening of ferroelectric polarization. Since the theoretical consideration of mismatch at curved surfaces is a complex separate problem, we leave such rigorous calculations for future; but consider the surface tension effect [31-34], which is also referred to as intrinsic surface stress for solids [56].

The paper has the following structure. Problem statement, containing free energy and basic equations with boundary conditions, is formulated in **Section II**. Shell models are described in **Subsection III.A. Subsections III.B** and **III.C** contain FEM results of the temperature evolution of domain structure and electric potential in $Sn_2P_2S_6$ and $BaTiO_3$ nanoparticles covered with shells, which dielectric permittivity is isotropic and temperature-independent, or tunable and temperature-dependent, isotropic or highly anisotropic. Analytical and numerical calculations of the $Sn_2P_2S_6$ and $BaTiO_3$ nanoparticles phase diagrams for various types of shells are presented in the end of **subsections III.B** and **III.C**, respectively. **Section IV** is a brief discussion with conclusive remarks. Euler-Lagrange equations, material parameters, and calculation details of the transition temperatures are presented in **Appendixes A, B** and **C,** respectively.

## II. PROBLEM STATEMENT

Let us consider a ferroelectric nanoparticle core of radius $R$ with a three-component ferroelectric polarization $\mathbf{P}(\mathbf{r})$ directed along one of the crystallographic axes. The core is regarded insulating, without any free charges. It is covered with a semiconducting shell of thickness $\Delta R$ that is characterized by relative dielectric permittivity tensor $\varepsilon_{ij}^{S}$. The particle in the shell is placed in the dielectric medium (polymer, gas, liquid, air, or vacuum) with "effective" dielectric permittivity, $\boldsymbol{\varepsilon}_e$. The word "effective" implies the presence of other particles in the medium, which can be described in effective medium approach. For the sake of clarity we regard that the medium is isotropic and temperature-independent, i.e. $\varepsilon_{ij}^{e} = \delta_{ij}\varepsilon_e$, in contrast to anisotropic and/or tunable shells. The considered physical model corresponds to a nanocomposite "core-shell nanoparticles in a dielectric medium" with a small volume fraction of ferroelectric nanoparticles (less than 10%) in the composite. The core-shell geometry is shown in **Fig. 1.**

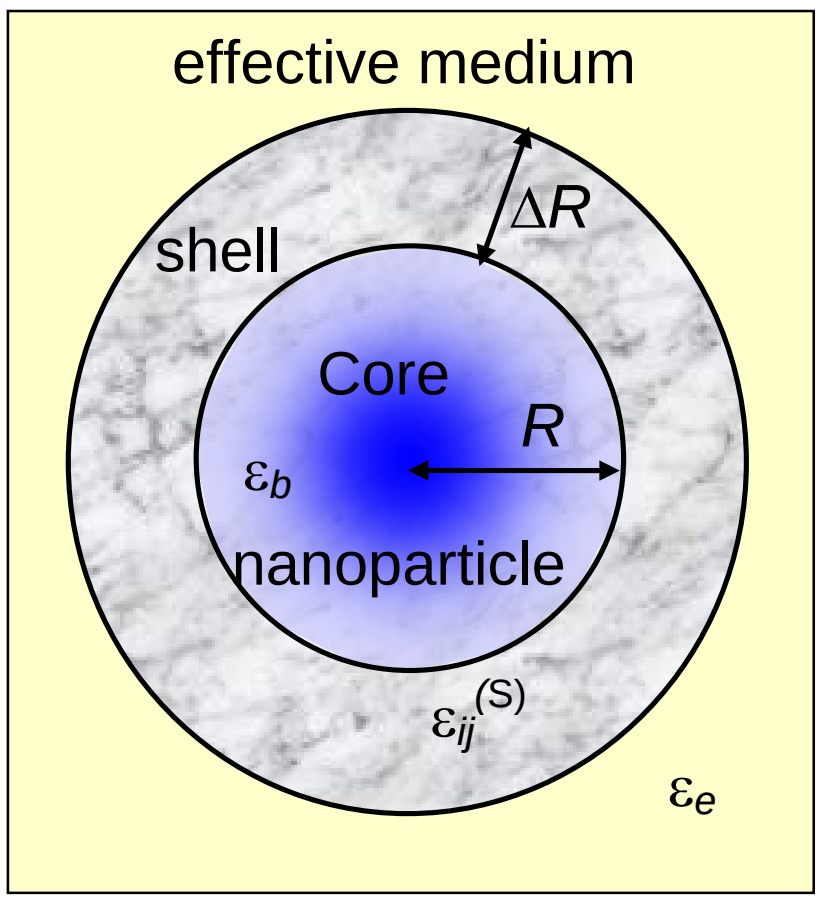

**FIGURE 1.** A spherical ferroelectric nanoparticle (core) covered with an isotropic or anisotropic semiconducting layer (shell) placed in an isotropic dielectric effective medium.

Below we assume that the shell is soft enough not to affect the strain and stress in the ferroelectric core. Thus the shell role is to modify and affect the electrostatics only, and so the elastic part of the problem is the same as in our previous study [39] of a ferroelectric core alone.

Since the ferroelectric polarization contains background and soft mode contributions, the electric displacement vector has the form $\mathbf{D}=\varepsilon_0\varepsilon_b\mathbf{E}+\mathbf{P}$ inside the particle, where $\varepsilon_b$ is a relative permittivity of the background material [57], and $\varepsilon_0$ is a universal dielectric constant. $D_i=\varepsilon_0\varepsilon_{ij}^S E_j$ in the shell and $D_i=\varepsilon_0\varepsilon^e E_i$ in the isotropic effective medium.

Electric field components $E_i$ are related to the electric potential $\varphi$ in a conventional way, $E_i=-\partial\varphi/\partial x_i$ . The potential $\varphi$ satisfies the Poisson equation in the ferroelectric core (subscript "$f$"):

$$\varepsilon_0\varepsilon_b\left(\frac{\partial^2}{\partial x_1^2}+\frac{\partial^2}{\partial x_2^2}+\frac{\partial^2}{\partial x_3^2}\right)\varphi_f=\frac{\partial P_i}{\partial x_i},\quad 0\le r\le R, \tag{1a}$$

and Debye-Hukkel equation [58] in the shell (subscript "s"):

$$\varepsilon_0\frac{\partial}{\partial x_i}\left(\varepsilon_{ij}^S\frac{\partial\varphi_s}{\partial x_j}\right)=-\varepsilon_0\frac{\varphi_s}{R_d^2},\qquad R<r<R+\Delta R, \tag{1b}$$

where $R_d$ is the "net" screening length of the shell. As mentioned in the introduction, there are several possible ways to change the dielectric permittivity tensor of a shell [49, 50, 52]. In all these cases the external stimuli (electric field, temperature, light, heating, etc.) affect the dielectric properties of the shell, which can influence the spatial distribution of ferro-active ions inside the core via electrostatic interactions and electric boundary conditions.

Outside the shell φ satisfies the Laplace equation:

$$\varepsilon_0\varepsilon_e\left(\frac{\partial^2}{\partial x_1^2}+\frac{\partial^2}{\partial x_2^2}+\frac{\partial^2}{\partial x_3^2}\right)\varphi_e=0,\quad r>R+\Delta R, \tag{1c}$$

Equations (1) are supplemented with the continuity conditions for electric potential and normal components of the electric displacements at the particle surface and core-shell interface:

$$\left(\varphi_e-\varphi_s\right)\Big|_{r=R+\Delta R}=0,\quad \mathbf{n}\left(\mathbf{D}_e-\mathbf{D}_s\right)\Big|_{r=R+\Delta R}=0,\quad \left(\varphi_S-\varphi_f\right)\Big|_{r=R}=0,\quad \mathbf{n}\left(\mathbf{D}_s-\mathbf{D}_f\right)\Big|_{r=R}=0. \tag{1d}$$

Since we do not apply an external field the potential vanishes either at infinity, $\varphi_e|_{r\to\infty}=0$, or at the surface of remote electrodes located at the boundaries of the computation cell.

LGD free energy functional $G$ additively includes Landau expansion on polarization 2-4-6 powers, $G_{Landau}$; polarization gradient energy contribution, $G_{grad}$; electrostatic contribution, $G_{el}$; elastic, electrostriction, and flexoelectric contributions, $G_{es+flexo}$. It has the form [30, 38, 39]:

$$G=G_{Landau}+G_{grad}+G_{el}+G_{es+flexo}, \tag{2a}$$

$$G_{Landau}=\int_{0<r<R}d^3r\left[\alpha P_i^2+a_{ij}P_i^2P_j^2+a_{ijk}P_i^2P_j^2P_k^2\right], \tag{2b}$$

$$G_{grad}=\int_{0<r<R}d^3r\frac{g_{ijkl}}{2}\frac{\partial P_i}{\partial x_j}\frac{\partial P_k}{\partial x_l}, \tag{2c}$$

$$G_{el}=-\int_{0<r<R}d^3r\left(P_iE_i+\frac{\varepsilon_0\varepsilon_b}{2}E_iE_i\right)-\frac{\varepsilon_0}{2}\int_{R<r<R+\Delta R}\varepsilon_{ij}^S E_iE_jd^3r-\frac{\varepsilon_0}{2}\int_{r>R+\Delta R}\varepsilon_{ij}^e E_iE_jd^3r, \tag{2d}$$

$$G_{es+flexo}=-\int_{0<r<R}d^3r\left(\frac{s_{ijkl}}{2}\sigma_{ij}\sigma_{kl}+Q_{ijkl}\sigma_{ij}P_kP_l+\frac{F_{ijkl}}{2}\left(\sigma_{ij}\frac{\partial P_l}{\partial x_k}-P_l\frac{\partial\sigma_{ij}}{\partial x_k}\right)\right)-\int_{R<r<R+\Delta R}d^3r\frac{s_{ijkl}^S}{2}\sigma_{ij}\sigma_{kl}. \tag{2e}$$

The coefficient $\alpha$ linearly depends on temperature *T:*

$$\alpha(T)=\alpha_T\left[T(R)-T_C^*\right], \tag{3a}$$

where $\alpha_T$ is the inverse Curie-Weiss constant and $T_C^*(R)$ is the Curie temperature renormalized by electrostriction and surface tension. Actually, the surface tension induces additional surface stresses $\sigma_{ij}$ proportional to the surface tension coefficient μ and equal to $\sigma_{11}=\sigma_{22}=\sigma_{33}|_{r=R}=-\frac{2\mu}{R}$ for a spherical nanoparticle of radius *R*. The stresses affect the Curie temperature and ferroelectric polarization behaviour due to the electrostriction coupling [29, 31-34]. Thus the renormalized Curie temperature, $T_C^*(R)$, acquires the following form [32, 39]:

$$T_C^*(R)=T_C\left(1-\frac{Q}{\alpha_T T_C}\frac{2\mu}{R}\right), \tag{3b}$$

where $T_C$ is a Curie temperature of a bulk ferroelectric. $Q$ is the sum of electrostriction tensor diagonal components that is positive for the most of ferroelectric perovskites with cubic m3m symmetry in the paraelectric phase, namely $0.004 < Q < 0.04\,\mathrm{m^4/C^2}$ [39]. Recent experiments tell us that μ is relatively small, not more than (2 – 4) N/m for most perovskites.

Tensor components $a_{ij}$ and $a_{ijk}$ are regarded temperature-independent. Tensor $a_{ij}$ is positively defined if the ferroelectric material undergoes a second order transition to the paraelectric phase and negative otherwise. Higher nonlinear tensor $a_{ijk}$ and gradient coefficients tensor $g_{ijkl}$ are positively defined and regarded temperature independent. The value $\sigma_{ij}$ is the stress tensor and $s_{ijkl}$ are elastic compliances tensor, $Q_{ijkl}$ is electrostriction tensor and $F_{ijkl}$ is the flexoelectric tenor in Eq.(2e).

Allowing for the Khalatnikov mechanism of polarization relaxation [59], minimization of the free energy (2) with respect to polarization leads to three coupled time-dependent Euler-Lagrange equations for polarization components,

$$\frac{\delta G}{\delta P_i} = -\Gamma \frac{\partial P_i}{\partial t}, \qquad (4)$$

where the explicit form for a ferroelectric nanoparticle with m3m parent symmetry is given in **Appendix A**. The boundary condition for polarization at the core-shell interface $r = R$ is natural and accounts for the flexoelectric effect:

$$\left. \left( g_{ijkl} \frac{\partial P_k}{\partial x_l} - F_{klij} \sigma_{kl} \right) n_j \right|_{r=R} = 0 \qquad (5)$$

where **n** is the outer normal to the surface, $i$=1, 2, 3.

Elastic stresses satisfy the equation of mechanical equilibrium in the nanoparticle and its shell,

$$\frac{\partial \sigma_{ij}}{\partial x_j} = 0, \qquad 0 < r < R + \Delta R. \qquad (6a)$$

Equations of state follow from the variation of the energy (2e) with respect to elastic stress, $\frac{\delta G_{es+flexo}}{\delta \sigma_{ij}} = -u_{ij}$, namely:

$$s_{ijkl}\sigma_{ij} + Q_{ijkl} P_k P_l + F_{ijkl} \frac{\partial P_l}{\partial x_k} = -u_{ij}, \qquad 0 < r < R \qquad (6b)$$

$$s^S_{ijkl}\sigma_{ij} = -u_{ij}, \qquad R < r < R + \Delta R \qquad (6b)$$

where $u_{ij}$ is the strain tensor. We will assume that $s^S_{ijkl} \approx s_{ijkl}$ for a "soft" shell. Elastic boundary conditions at the particle core-shell interface $r = R + \Delta R$ are the continuity of the elastic displacement vector and normal stresses.

## III. RESULTS AND DISCUSSION

### A. Shell models

Below we consider three typical varieties of the shell dielectric properties, namely a polymer shell with temperature-independent isotropic dielectric permittivity (a), tunable paraelectric shell with isotropic strongly temperature-dependent dielectric permittivity (b), and anisotropic tunable shell of liquid crystal (**LC**) (c), with a temperature dependent anisotropic dielectric permittivity. Mathematical expressions for the shells permittivity are given below.

**(a)** A polymer (or glass) shell has the isotropic temperature-independent dielectric permittivity

$$\varepsilon_{11}^{S}=\varepsilon_{22}^{S}=\varepsilon_{33}^{S}=\varepsilon_{P}^{S}. \tag{7a}$$

The values of $\varepsilon_{P}^{S}$ from 3 to 15 correspond to inorganic glasses (e.g. PMMA [60]) or organic polymers (e.g. PVDF [61]).

**(b)** A tunable shell of paraelectric strontium titanate ($SrTiO_3$) has an isotropic strongly temperature-dependent dielectric permittivity,

$$\varepsilon_{11}^{S}=\varepsilon_{22}^{S}=\varepsilon_{33}^{S}=\varepsilon_{PE}^{S}(T)=\frac{1}{\varepsilon_0\alpha_T T_q^{(E)}}\left(\coth\left(\frac{T_q^{(E)}}{T}\right)-\coth\left(\frac{T_q^{(E)}}{T_0^{(E)}}\right)\right)^{-1}, \tag{7b}$$

with the Curie-Weiss parameter $\alpha_T$ =0.75×10$^6$ m/(F K), and characteristic temperatures $T_0^{(E)}=30$ K and $T_q^{(E)}=54$ K [62]. It should be noted that $\varepsilon_{PE}^{S}(T)\approx 3000$ at $T$=50 K and $\varepsilon_{PE}^{S}(T)\approx 300$ at T=293 K, allowing the spontaneous polarization of the ferroelectric core to be effectively screened by the tunable shell at room and lower temperatures.

**(c)** An anisotropic tunable LC shell can have a strongly temperature dependent anisotropic dielectric permittivity. Below we consider the temperature dependence of $\varepsilon_{ij}^{S}$ taken from Ref.[51],

$$\varepsilon_{\perp}^{S}=\text{const},\qquad \varepsilon_{\parallel}^{S}=\varepsilon_{\perp}^{S}+\begin{cases}0, & T>T_{cn},\\ \Delta\varepsilon\left(1-\frac{T}{T_{cn}}\right)^{\beta}\exp\left(\frac{E_a}{k_B T}\right), & T_m\le T<T_{cn},\\ \Delta\varepsilon\left(1-\frac{T_m}{T_{cn}}\right)^{\beta}\exp\left(\frac{E_a}{k_B T_m}\right), & T<T_m.\end{cases} \tag{7c}$$

To study the anisotropy effect, the temperature of the shell transition to isotropic state, $T_{cn}$, should be smaller than the temperature $T_C^{*}(R)$. From Eq.(7c), dielectric constants of the semiconducting LC shell are tunable and critically depend on temperature. For parameters $\varepsilon_{\perp}^{S}=3$, $\Delta\varepsilon=0.0245$, $\beta$=0.5, $T_{cn} = 341$ K, $E_a = 0.281$ eV, and $T_m \approx 271$ K, taken from Ref.[51], we determined that the shell anisotropy vanishes in the "polar" direction z at temperatures $T > T_{cn}$. Simultaneously, the permittivity value decreases sharply in this direction diminishing the screening of the nanoparticle ferroelectric polarization. For temperatures $T << T_{cn}$ the shell is strongly dielectrically anisotropic, $\varepsilon_{\|}^{S} >> \varepsilon_{\perp}^{S}$, e.g. $\varepsilon_{\perp}^{S} = 3$ and $\varepsilon_{\|}^{S} = 629$ at room temperature $T$=293 K, while $\varepsilon_{\|}^{S} \cong 1856$ at $T$=271 K. Note that the second line in Eq.(7c) cannot be valid below the LC melting temperature $T_m$, but as a "simple model" we will assume that the anisotropic shell acts as a "static anisotropic semiconductor" in the direction of LC director at temperatures $T << T_m < T_{cn}$. The case of the temperature dependence of $\varepsilon_{\|}^{S}$ for $T < T_m$ can be described by the third line in Eq.(7c).

Since further calculations will go well beyond the temperature range of conventional LC phase, and we will consider (5-10) nm thick shells, which is about 2-5 LC molecules, one may argue that a typical LC phase can lose its bulk properties at this scale. Allowing for the fact, we will consider a shell made of a hypothetic "LC-like material" instead of "true" LC shell. However, for the sake of brevity, we will use the abbreviation "**LC shell**" hereinafter, instead of the longer expression "shell of LC-like material".

### B. "Tunable" domains in uniaxial ferroelectric nanoparticles

FEM has been performed to find the solution of a coupled system (1)-(6) for the case of a uniaxial ferroelectric $Sn_2P_2S_6$ nanoparticle covered with a semiconducting shell having various dielectric properties given by Eqs.(7). The Curie temperature of bulk ferroelectric $Sn_2P_2S_6$ is 336 K; other LGD parameters, collected from Refs.[63, 64, 65] and references therein, are listed in **Table AI, Appendix A.** The spontaneous polarization $P_3$ is directed along the polar axis 3.

Shell tuning can help to control the temperature evolution of domain structure, which is formed in a $Sn_2P_2S_6$ nanoparticle of radius of (2–10) nm from an initial randomly small distribution of polarization (commonly called "random seeding"). Nanoparticles with a radius of less than 2 nm were not considered due to the fact that the phenomenological LGD approach is not applicable for the sizes below 10 lattice constants [28-39]. Core-shell particles are placed in a dielectrically isotropic ambience with $\varepsilon_e = 10$. An external field is absent.

Simulation results for the particles of 10-nm radius covered with either isotropic temperature-independent high-k dielectric polymer [model Eq.(7a)], or tunable highly polarizable paraelectric $SrTiO_3$ [model Eq.(7b)] are shown in **Figs. 2** and **3,** respectively. Simulation results for the particles covered with dielectrically anisotropic LC tunable shells [model Eq.(7c)] with parallel and perpendicular orientation of the anisotropy axis with respect to the particle polar axis 3 are shown in **Figs. 4** and **5,** respectively. Distributions of $P_3(x,y)$ in the equatorial XY-section at z=0 are shown in plots **(a).** Distributions of $P_3(x,z)$ and $\varphi(x,z)$ in the polar XZ-section at y=0 are shown in plots **(b)** and **(c)**, respectively. Different plots correspond to the temperature values $T$=(50 – 300) K indicated in the graphs. Potential $\varphi$ is zero at z=0. Only a half-particle is shown in **Figs. 2-5(b-c)**, because all distributions in the other half are mirror-symmetric in XZ cross-sections.

Analyzing **Figs.2-5**, we see that ferroelectric nanoparticles covered with a tunable shell have several phases, namely, single-domain ferroelectric (**SDFE**) phase and poly-domain ferroelectric (**PDFE**) states, including stable labyrinthine domains (**LD**), and/or meandering domain (**MD**) walls, and paraelectric (**PE**) phase. Note, we used various initial distributions to check that the LD and/or MD arising from the random seeding correspond to the minimal energy at a given temperature. We compared energies in all sensitive cases. For instance it appeared that the configuration when the shell and core polar axes coincide has higher energy than the configuration when they are perpendicular. Comparing the temperature behavior of the domain structure in the particles of the same radius (10 nm) covered with shells of the same thickness (5 nm) and with the same screening length (2 nm), but with different dielectric properties, the following trends are revealed.

For a polymer shell with a temperature-independent, but relatively high isotropic dielectric permittivity ($\varepsilon_P^S = 15$), the SDFE state of the particle is absent [see **Figs. 2(a,b)**]. Even at low temperatures, the LD structure is the most stable, its small size (about 5 lattice constants) borders with the spatially modulated phase. As the temperature increases, the polarization amplitude gradually decreases, so the contrast of LD slowly decreases (they "fade"), and simultaneously the labyrinthine pattern slowly turns into curved stripes surrounded by the PE layer near the particle surface. In this case the size of the domain stripes practically does not change from 50 K to 200 K, when the particle all is almost filled with the PE region from the surface to the center. The alternating electric potential is maximal near the poles of the particle, and the broadening of the domain walls occurs there, which leads to a decrease in the depolarization electric field [see **Figs. 2(c)**]. With increasing temperature, the potential decreases and tends to zero below PE phase transition, which is slightly above 200 K. Note that the reduced screening length, $R_d / \varepsilon_P^S =$ 0.13 nm, is much smaller than $R_d$ due to the dielectric screening effect.

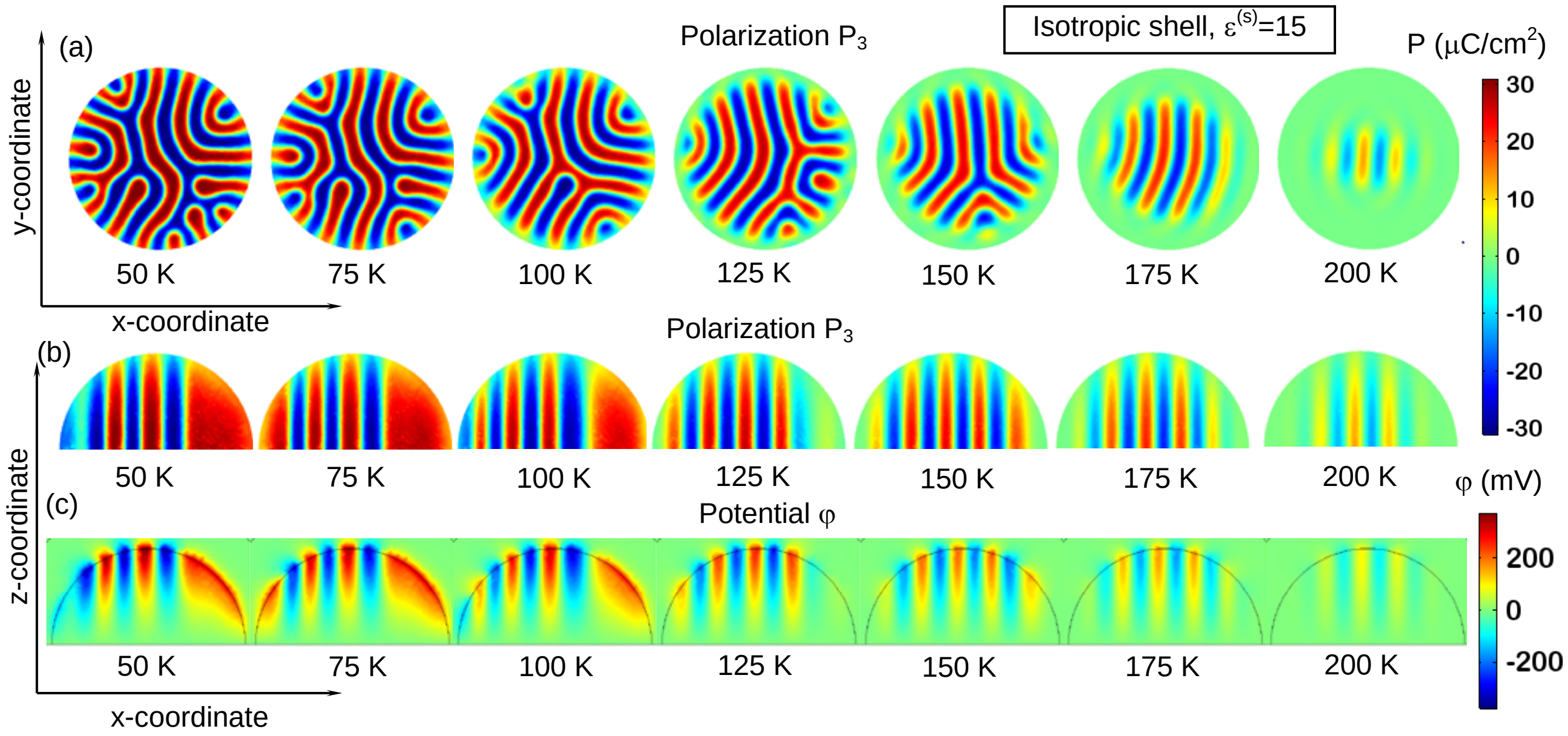

**FIGURE 2.** Polarization and electrostatic potential distributions, $P_3(x,y,z)$ and $\varphi(x,y,z)$, inside a $Sn_2P_2S_6$ nanoparticle covered with a polymer shell having isotropic temperature-independent dielectric permittivity $\varepsilon_P^S = 15$. **(a)** Distribution of $P_3(x,y)$ in the XY-section at z = 0**.** Distributions of $P_3(x,z)$ **(b)** and $\varphi(x,z)$ **(c)** in the XZ-section at $y$ = 0. Different plots corresponds to the temperatures $T$ = 50, 75, 100, 125, 150, 175, and 200 K indicated under the graphs. Particle radius $R$ = 10 nm, shell thickness $\Delta R$ = 5 nm and effective screening length $R_d = 2$ nm.

For a tunable $SrTiO_3$ shell, where the isotropic permittivity is high (about 300 at 293 K) and sharply increasing with decreasing temperature (about 3000 at 50 K), the SDFE state of the particle is stable up to 180 K [see **Figs. 3(a,b)**]. At higher temperatures, the LD-like and MD-type structures are stable, and their characteristic size (about 15 lattice constants) is significantly higher than in the case of polymer shell, shown in **Figs. 2(a)**. With an increase in temperature from 200 K to 280 K, the amplitude of polarization gradually decreases, and the contrast of domain pattern fades, domains slightly change in shape and size, and their meandering walls becomes significantly wider. Above 280 K, the domain structure disappears, since the particle becomes paraelectric. The electric potential is maximal in the shell and near the particle poles, with the exception of the domain walls, which slightly broaden when approaching the surface [see **Figs. 3(c)**]. With increasing temperature, the potential value gradually decreases and tends to zero in the PE phase. Note that the reduced screening length, $R_d/\varepsilon_{PE}^S$, changes from 0.007 nm at 50 K to 0.067 nm at 293 K, which is much smaller than in the case of a polymer shell. Comparing **Figs. 2** and **3** we conclude that a temperature-tunable paraelectric shell provides much more efficient screening of the nanoparticle polarization in comparison with temperature-independent polymer shell.

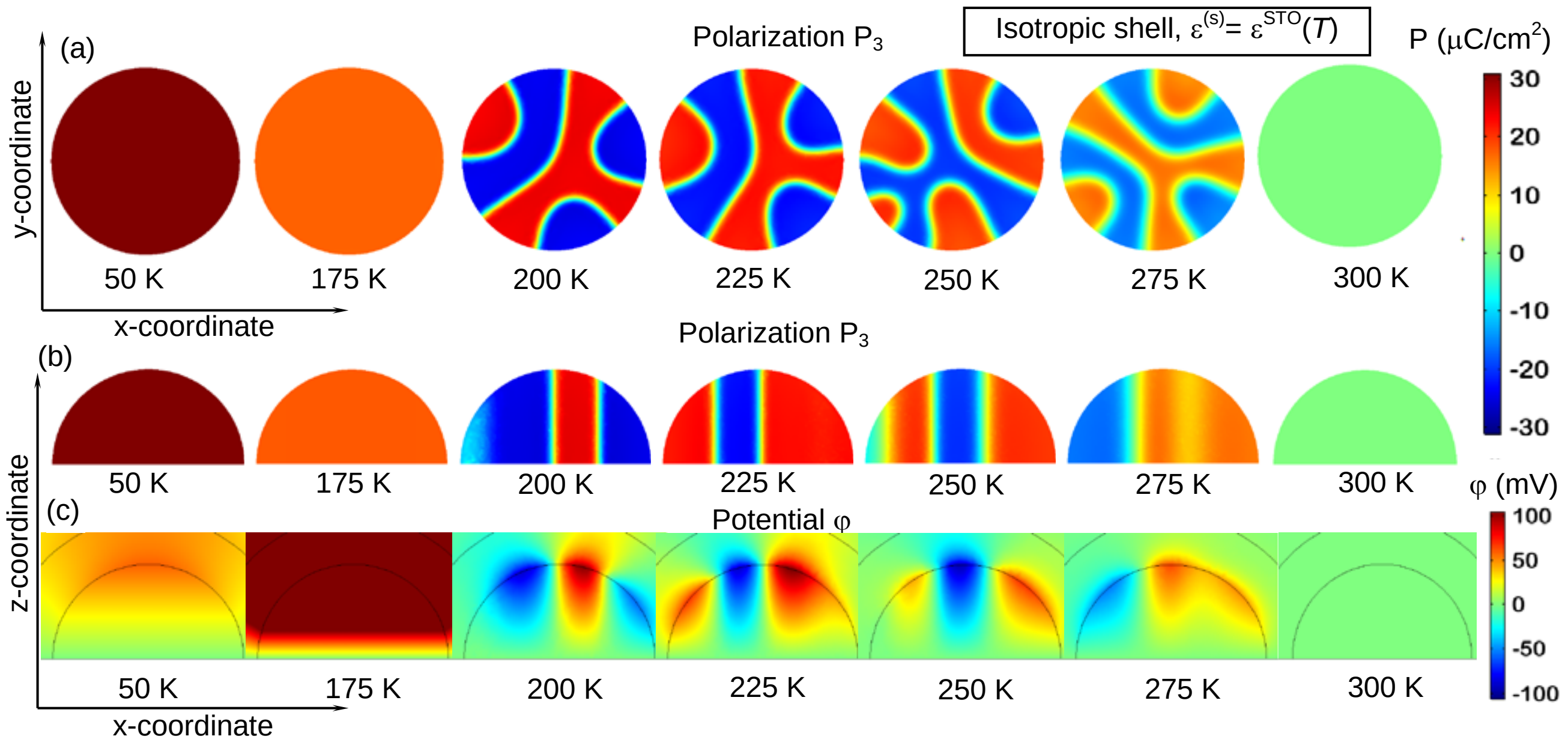

**FIGURE 3.** Polarization and electrostatic potential distributions, $P_3(x,y,z)$ and $\varphi(x,y,z)$, inside a $Sn_2P_2S_6$ nanoparticle covered with a $SrTiO_3$ shell with isotropic temperature-dependent dielectric permittivity given by Eq.(7b). The distribution of $P_3(x,y)$ in the XY-section at $z$ = 0 is shown in plot **(a).** Distributions of $P_3(x,z)$ and $\varphi(x,z)$ in the XZ-section at $y$ = 0 are shown in plots **(b)** and **(c)**, respectively. Different plots correspond to the temperatures $T$ = 50, 175, 200, 225, 250, 275, and 300 K indicated under the graphs. Parameters $R$, $\Delta R$, and $R_d$ are the same as in **Fig. 2.**

For a tunable LC shell that has an anisotropic dielectric permittivity described by Eq.(7c) and sharply decreases with increasing temperature, two principally different cases are possible depending on the mutual orientation of the shell anisotropy axis and the polar axis of the nanoparticle.

For the parallel orientation, shown in **Figs. 4**, the stable SDFE state of the particle is absent down to 0 K, instead PDFE state exists from 0 K up to 225 K. The PDFE state represents itself nested ring-shaped cylindrical domains, where the number of rings slightly increases with increasing temperature [see **Fig. 4(a)**]. The nested domains lose axial symmetry due to the meandering of the domain walls. Simultaneously with the increase in the number of ring-shaped MD, the polarization amplitude gradually decreases, the domain pattern contrast fades, and the domain walls become thicker and wider when approaching the surface [see **Fig. 4(a)-(b)**]. Above 230 K, the domain structure disappears, since the particle becomes PE. The sign-alternating electric potential is non-zero throughout a significant part of the particle, and reaches maximal values in the shell [see **Figs. 4(c)**]. With increasing temperature, the potential value and modulation depth gradually decrease and become zero in the PE phase.

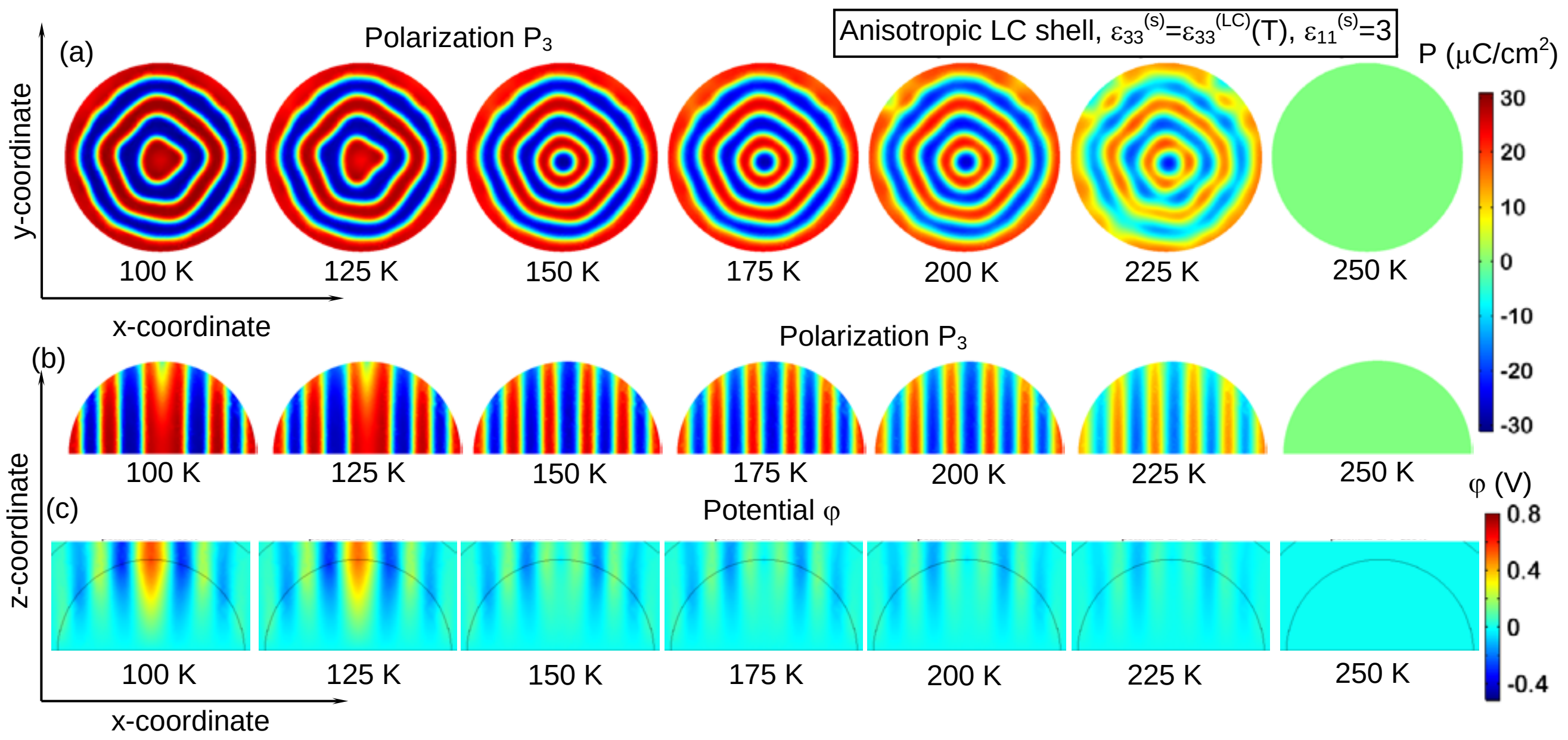

**FIGURE 4.** Polarization and electrostatic potential distributions, $P_3(x,y,z)$ and $\varphi(x,y,z)$, inside a $Sn_2P_2S_6$ nanoparticle covered with a tunable LC shell with an anisotropic temperature-dependent dielectric permittivity given by Eq.(7b). The shell anisotropy axis 3 is parallel to the particle polarization $P_3$. The distribution $P_3(x,y)$, in the XY-section at $z = 0$ is shown in plot **(a).** Distributions of $P_3(x,z)$ and $\varphi(x,z)$ in the YZ-section at $x$=0 are shown in plots **(b)** and **(c)**, respectively. Different plots correspond to the temperatures $T$= 100, 125, 150, 175, 200, 225, and 250 K indicated under the graphs. Parameters $R$, $\Delta R$, and $R_d$ are the same as in **Fig. 2.**

For the perpendicular orientation shown in **Fig. 5**, the SDFE state of a particle is also unstable for all temperatures, instead the stable PDFE state exists and transforms into slightly curved domain stripes with increasing temperature up to 260 K [see **Fig. 5(a)**]. The wall curvature and meandering slightly increases, the polarization amplitude gradually decreases, and the domain pattern contrast fades as the temperature increases. Also the domain walls become thicker and wider as they approach the surface [see **Fig. 5(b)**]. Close to the PE transition, non-poled regions occur in the equatorial plane. Above 285 K, the domain structure disappears, since the particle becomes PE. The sign-alternating electric potential, shown in **Fig. 5(c)**, is weak but non-zero in a small part of the particle. The potential maxima are located in the shell and are small compared to the values shown in **Fig. 4**. With increasing temperature, the potential gradually decreases and tends to zero in the PE phase.

Unexpectedly the configuration where the shell and core polar axes are parallel has a greater energy than the perpendicular configuration.

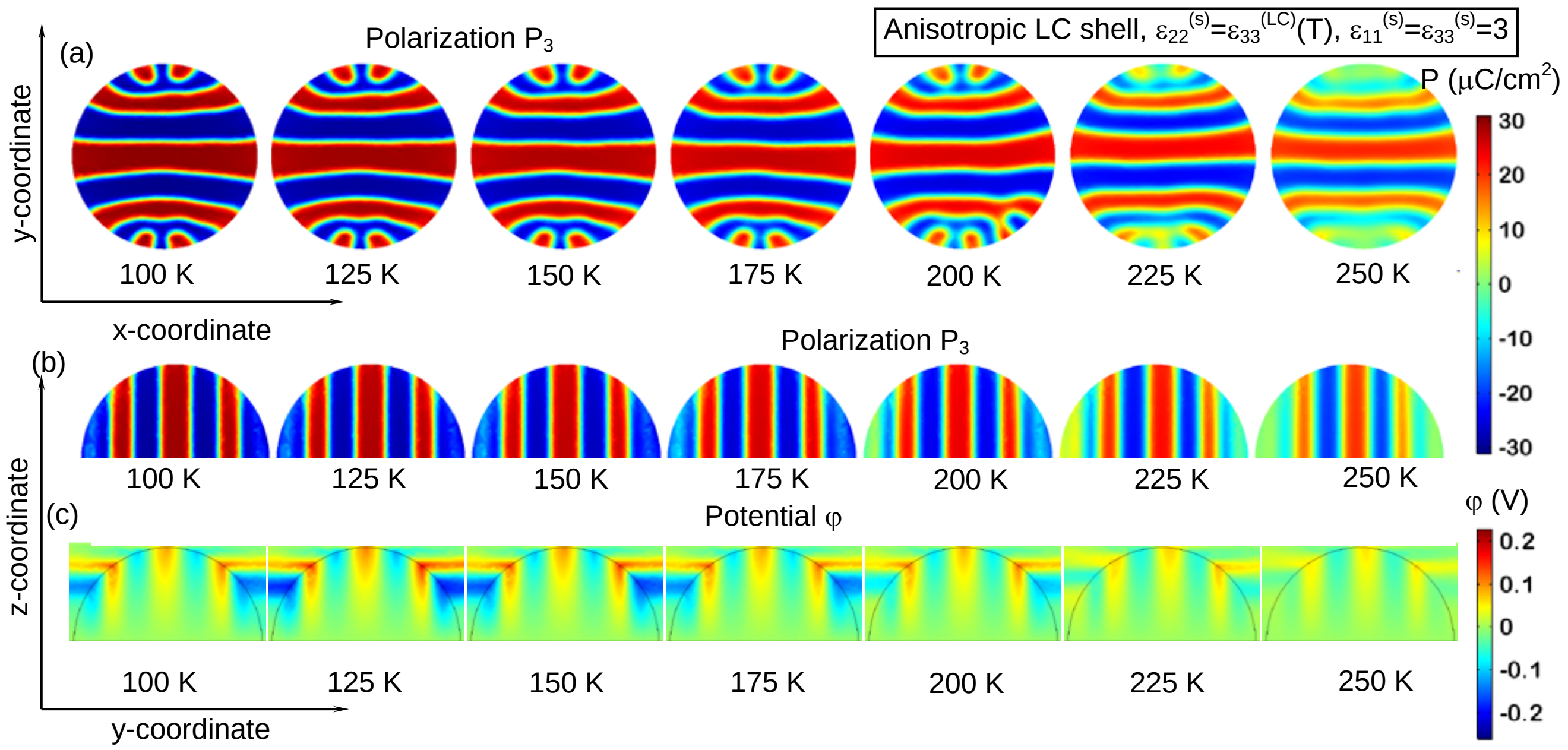

**FIGURE 5.** Polarization and electrostatic potential distributions, $P_3(x,y,z)$ and $\varphi(x,y,z)$, inside a $Sn_2P_2S_6$ nanoparticle covered with a tunable LC shell with an anisotropic temperature-dependent dielectric permittivity given by Eq.(7c). The shell anisotropy axis 3 is perpendicular to the particle polarization $P_3$. The distribution of $P_3(x,y)$ in the equatorial XY-section at $z = 0$ is shown in plot **(a).** Distributions of $P_3(x,z)$ and $\varphi(x,z)$ in the polar XZ-section at $y = 0$ are shown in plots **(b)** and **(c)**, respectively. Different plots correspond to the temperatures $T$= 100, 125, 150, 175, 200, 225, and 250 K indicated under the graphs. Parameters $R$, $\Delta R$, and $R_d$ are the same as in **Fig. 2.**

Comparing **Figs. 4-5** with **Figs. 2-3** we conclude that a temperature-tunable anisotropic LC shell provides much more efficient screening of the nanoparticle polarization in comparison to the temperature-independent polymer shell. The tunable anisotropy adds a new level of functionality for the control of the ferroelectric domain structure in comparison with an isotropic paraelectric shell. Note, that the temperatures T = (50 - 280) K characteristic to the relatively deep ferroelectric phase of $Sn_2P_2S_6$ are low for typical LCs. However, the calculations show the fundamental possibilities for controlling the LD and MD morphologies in other ferroelectrics with e.g. higher $T_C$, which make possibleto consider them as advanced cryptographic materials [66].

Since there are real possibilities to tune the value and anisotropy of the shell dielectric permittivity, FEM results, shown in **Figs. 2-5,** indicate the opportunities to control the domain structure of uniaxial ferroelectric nanoparticles covered with a tunable shell. To support the idea on a prognostic level, analytical calculations are necessary.

As it can be seen from **Figs. 2-5**, a spherical nanoparticle of fixed radius $R$ is in the PE phase at higher temperatures, or in the SDFE or PDFE states at lower ones. The transition temperature between various domain states depends on the dielectric properties of the tunable shell. Actually, an approximate analytical expression for the transition temperature from SDFE to PE phase is:

$$T_{SDFE}(R)\approx T_C^*(R)-\frac{1}{\alpha_T\varepsilon_0\left[\varepsilon_b+\varepsilon_{eff}^S\left(2+\frac{R^2}{R_d(R_d+R)}\right)\right]},\tag{8}$$

where the temperature $T_C^*(R)$ is given by Eq.(3b) and $\varepsilon_{eff}^S$ is the effective dielectric permittivity of the shell. $\varepsilon_{eff}^S$ coincides with $\varepsilon_P^S$ for an isotropic polymer, and with $\varepsilon_{STO}^S(T)$ for an isotropic paraelectric; for the case of anisotropic LC shell it coincides with $\frac{\varepsilon_{||}(T)}{4}\left(1+\sqrt{1+8\frac{\varepsilon_\perp}{\varepsilon_{||}(T)}}\right)$. The second term in Eq.(8) originates from a depolarization field [31]. Note that expression (8) becomes a transcendental equation for the temperature dependent $\varepsilon_{eff}^S(T)$.

The derivation of Eq.(8) is given in **Appendix B**, where we supposed that the polarization gradient inside the core is small, the core-shell system is relatively "thick" and/or the shell is "conductive" enough, i.e. $(R+\Delta R)>>R_d$, and/or $\left|\varepsilon_e-\varepsilon_{eff}^S\right|<<\varepsilon_{eff}^S$. Under these assumptions the external electric field exponentially vanishes inside the particle, $E_i^{ext}\approx\frac{6\varepsilon_e\exp(-\Delta R/R_d)}{(\varepsilon_b+2\varepsilon_{eff}^S)R_d+\varepsilon_{eff}^S R}E_{ext}$. More rigorous expressions, which are valid for arbitrary $R, \Delta R, R_d$, and $\varepsilon_e$ are given by the cumbersome Eqs.(B.8). Note that Eq.(8) does not account for the shell anisotropy in a rigorous manner, and the approximate expression for $\varepsilon_{eff}^S(\varepsilon_\perp,\varepsilon_{||})$ is derived in subsection B3 of **Appendix B**.

An approximate analytical expression for the nanoparticle transition temperature from PDFE state to PE phase has the form:

$$T_{PDFE}(R)\approx T_C^*(R)-\frac{1}{\alpha_T}\left(g_{44}k_{\min}^2+\frac{1}{\varepsilon_0}\left[\frac{\varepsilon_{eff}^S R^2}{R_d(R_d+R)}+(\varepsilon_b+2\varepsilon_{eff}^S)(1+k_{\min}^2(\xi R)^2)\right]^{-1}\right).\tag{9a}$$

Here the first term in the parenthesis originated from the correlation effect and the second term is from the depolarization field energy of the domain stripes. The wave vectors **k** corresponding to the domain structure onset can have any direction, since only their absolute value is fixed, $|\mathbf{k}|=k_{\min}$, where $k_{\min}$ is given by expression:

$$k_{\min}=\sqrt{\frac{2}{\sqrt{\varepsilon_0(\varepsilon_b+2\varepsilon_{eff}^S)g_{44}}R}-\left(1+\frac{\varepsilon_{eff}^S R^2}{(\varepsilon_b+2\varepsilon_{eff}^S)R_d(R_d+R)}\right)\left(\frac{1}{\xi R}\right)^2}.\tag{9b}$$

Parameter $\xi$ is a geometrical factor close to 1/4. The derivation of Eqs.(9) is given in **Appendix C.**

Expressions (9) have physical sense under the condition

$$\frac{1}{\sqrt{\varepsilon_0(\varepsilon_b+2\varepsilon_{eff}^S)g_{44}}}\geq\frac{1}{\xi R}+\frac{2\varepsilon_{eff}^S R}{(\varepsilon_b+2\varepsilon_{eff}^S)R_d(R_d+R)}\tag{10}$$

The fulfilment of the inequality in Eq.(10) corresponds to the single domain state transition that occurs in a tricritical point on the phase diagram, where the energies of SDFE state and PDFE phase are equal to the PE phase energy . In the tricritical point $k_{\min}=0$ and $T_{PDFE}(R)=T_{SDFE}(R)$. The radius dependence of the tricritical point temperature $T_{tcr}(R)$ and effective screening length $R_d^{trc}$ can be found as [38]:

$$T_{tcr}(R)=T_C^*(R)-\frac{\sqrt{\varepsilon_0\left(\varepsilon_b+2\varepsilon_{eff}^S\right)g_{44}}}{\alpha_T\varepsilon_0\left(\varepsilon_b+2\varepsilon_{eff}^S\right)\xi R}, \qquad (10a)$$

$$\frac{1}{R_d^{trc}}=\frac{\varepsilon_b+2\varepsilon_{eff}^S}{\varepsilon_{eff}^S}\left(\frac{\xi}{\sqrt{\varepsilon_0\left(\varepsilon_b+2\varepsilon_{eff}^S\right)g_{44}}}-\frac{1}{R}\right). \qquad (10b)$$

Note that if one considers the product $\varepsilon_{eff}^S$ to be temperature-dependent for temperature-tunable paraelectric and anisotropic shells, expressions (8)-(10) transform into transcendental equations.

Approximate analytical expression for the nanoparticle transition temperature from PDFE to SDFE states can be estimated from their free energy equality, since the transition is of the first order. Actually, the stability of a PDFE state in comparison with a SDFE state depends on the balance between the depolarization field energy (appearing from incomplete screening of the spontaneous polarization characterized by the effective screening length $R_d/\varepsilon_{eff}^S$ ) and the domain walls energy (proportional to the gradient coefficient $g_{44}$). The domain splitting starts when it becomes energetically preferable.

Phase diagrams of a $Sn_2P_2S_6$ nanoparticle covered with a semiconducting shell are shown in **Figs. 6,** where plots **(a)**-**(d)** correspond to various shells with the same thickness $\Delta R$=5 nm and effective screening length $R_d=2$ nm. The stable PE phase, SDFE, and PDFE (with or without MD and/or LD) states are separated by the boundaries (solid, dashed, and dotted curves) calculated from Eqs.(8)-(9). The curves' behavior has been checked by FEM. To obtain the best agreement with FEM results, the fitting parameter, dimensionless geometrical factor $\xi$, appeared equal to 0.25 for the non-tunable [plot **(a)**] and polymer [plot **(b)**] shells, 0.20 for the $SrTiO_3$ shell [plot **(c)**], and 0.3 for the LC shell [plot **(d)**]. Notably the meta-stability or stability of PDFE state with domain stripes ("striped" PDFE) against PDFE phase with MD and/or LD were determined from the comparison of the free energies corresponding to these domain morphologies. The PE phase energy is zero, as anticipated.

The diagram in **Fig. 6(a)** shows the typical changes of phases and domain morphologies, which happen in the 10-nm nanoparticle as a function of $T$ and $\varepsilon_{eff}^S$. A SDFE state is stable at rather high $\varepsilon_{eff}^S>(300-1000)$, which corresponds to lower temperatures. Two- or poly-domain stripes are stable at $30<\varepsilon_{eff}^S<300$, which corresponds to higher temperatures. Coexistence of striped PDFE, LD and MD morphologies appears at relatively small permittivity values, $\varepsilon_{eff}^{S} < 30$, and is followed by the size-induced phase transition into the stable PE phase at $T > T_{PDFE}\left(\varepsilon_{eff}^{S}\right)$. MD and LD are stable at $T$<200K, become metastable and coexist with domain stripes at the particle surface at higher temperatures (200 – 300) K, and transform into the PE phase at $T \geq 300$ K. PE, PDFE and SDFE coexist in the tricritical point, denoted by *tcr* (about 330 K and $R \approx 1$ μm)

The typical changes of phase diagrams of the nanoparticles taking place with increasing *T* and *R* are shown in **Fig. 6(b-d)** for various shells.

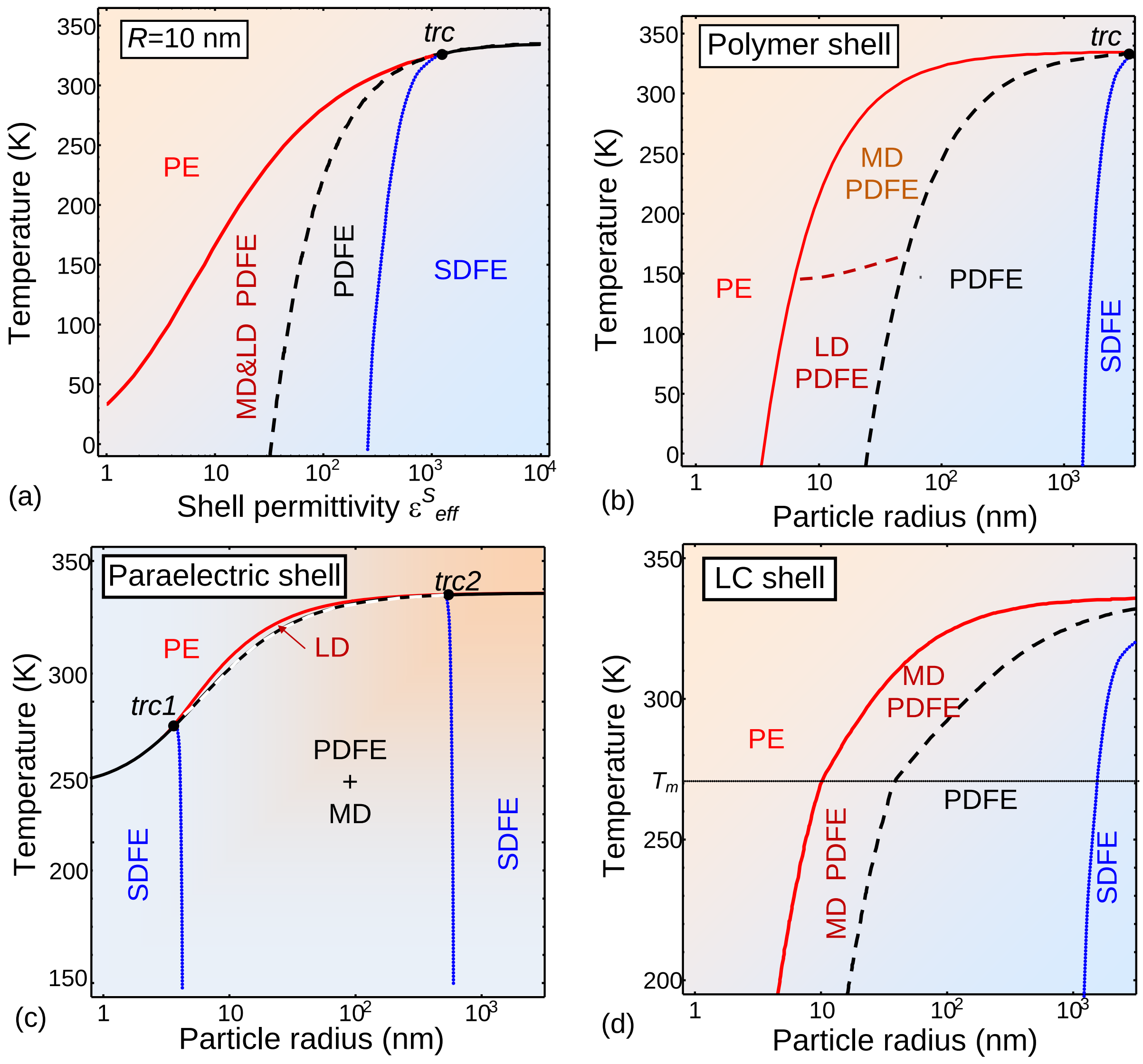

**FIGURE 6.** Phase diagrams of a $Sn_2P_2S_6$ nanoparticle covered with a semiconducting shell. **(a)** The diagram in coordinates "*T*" and "$\varepsilon_{eff}^{S}$"calculated for *R* = 10 nm, shell thickness Δ*R* = 5 nm, and effective screening length $R_d = 2$ nm. **(b)-(d)** Diagrams in coordinates "*T*" and "*R*" calculated for polymer shell with $\varepsilon_{eff}^{S} = 15$ **(b),** paraelectric $SrTiO_3$ shell **(c)**, and LC shell **(d)**. Stable ferroelectric single domain (SDFE), ferroelectric polydomain (PDFE) states, including labyrinthine (LD) and meandering (MD) domains, and paraelectric (PE) phases are shown. Red solid curves are PDFE-PE phase boundaries, blue dotted curves are SDFE-PDFE phase

boundaries, and dashed black curves separate the regions of PDFE phase with LD and/or MD domain morphology from the striped PDFE phase.

If the particle is covered with a polymer ($\varepsilon_{eff}^{S}=15$), its phase diagram has a standard form shown in **Fig. 6(b)**. At small radii, the particle is in a PE phase (the region above the solid red curve), which borders the PDFE phase, and the boundary between them shifts to smaller radii with decreasing temperature (the region left to the solid red curve). Near the boundary of the PE phase, the PDFE phase contains LD, which transform into MD with increasing temperature (the area between the solid red and dashed black curves). The width of the region containing LD and MD decreases with increasing temperature, and the region itself shifts toward larger radii. With a further increase in the radius, the particle transforms into striped PDFE (the region between the black dashed and dotted blue curves). For very large radii (above 1μm), the particle becomes single-domain (the region right of the blue dotted curve). PE, PDFE, and SDFE coexist in the tricritical point *tcr* at about 330 K and $R \approx 4$ μm.

If the particle is covered with the paraelectric $SrTiO_3$ shell, its phase diagram, shown in **Fig. 6(c)**, has a form significantly different from the diagram for a particle in the polymer shell [shown in **Fig. 6(b)**]. Due to the fact that the dielectric permittivity of $SrTiO_3$ increases significantly with decreasing temperature, there are two regions of SDFE phase located at small and large radii, between which there is a PDFE phase with a transitional morphology of the domain structure. Namely, there are domains with curved meandering walls, indicated in the diagram as “MD PDFE”, although the winding labyrinth is not formed despite the tendency of the domain walls branching (see **Fig. 3**). There is a narrow region of (meta)stable LD near the boundary between PE and MD PDFE phases. PE, PDFE, and SDFE phases coexist in two tricritical points, labeled “*tcr1*” and “*tcr2*”, which correspond to $R$~4 nm and $R$~500 nm, respectively.

If the particle is covered with a tunable anisotropic dielectric (e.g. LC), the phase diagram [shown in **Fig. 6(d)**] has a form similar to that shown in **Fig. 6(b)** for a particle in a polymer shell. Therefore, all comments made to **Fig. 6(b)** are qualitatively valid for **Fig. 6(d)**. However, there are a few differences. At the boundaries between PE and PDFE phases a break is observed at a temperature $T_m$ corresponding to the melting of “frozen” LC. The region containing LD and MD is narrower at $T<T_m$ than the one in **Fig. 6(b)**, and at $T>T_m$, the width of LD+MD area is larger in comparison with the one in **Fig. 6(b)**. The tricritical point is not reached even for micron-sized particles, although a tendency towards convergence of the PE, PDFE, and SDFE phase boundaries is observed.

**Figure 6** demonstrates the ability to control the phase state and domain structure of spherical uniaxial ferroelectric particles by selecting the tunable shell and particle radius. The most interesting results can be achieved for the particles covered with temperature-tunable shells.

### C. Closure domains and vortexes in multiaxial ferroelectric nanoparticles

Using FEM to find the solution of a coupled system (1)-(6), we simulated the domain structure, which is formed in a nanoparticle of multiaxial ferroelectric barium titanate ($BaTiO_3$) with a radius of (2–10) nm covered with a semiconducting shell with a constant or temperature-tunable dielectric permittivity tensor. Curie temperature of bulk $BaTiO_3$ is 381 K; other LGD parameters, collected from Refs.[67, 68, 69, 70, 71] and references therein, are listed in **Table AII, Appendix A.** Core-shell nanoparticles are placed in a dielectrically isotropic ambience with $\varepsilon_e = 10$.

Simulation results for the polarization distributions, which are final (i.e. "time-relaxed") state with minimal energy of a nanoparticle covered with isotropic temperature-independent high-k polymer [model Eq.(7a)], or tunable highly polarizable paraelectric $SrTiO_3$ [model Eq.(7b)] are shown in **Figs. 7** and **8,** respectively. Simulation results for the particle covered with a dielectrically anisotropic tunable LC shell [model Eq.(7c)] are shown in **Figs. 9**.

**Figures 7-9** illustrate the temperature evolution of $P_i$ in the equatorial XY- and polar XZ - sections of a nanoparticle. Different plots correspond to the different temperatures in the range (50 – 390) K indicated above the plots. White arrows in **Figs. 7-9** show the XY and XZ projections of the polarization vector **P**; while its components $P_i$ are shown by the color scale. Only a half-particle is shown in XZ-sections, because all distributions in the other half are mirror-symmetric.

There are rhombohedral, vortex-like orthorhombic, tetragonal, and paraelectric phases, which are absolutely stable at different temperatures. The temperature ranges of the phases stability depend slightly on the dielectric properties of the shell. The rhombohedral and vortex-like orthorhombic phases can coexist at some temperatures; the temperature range of the tetragonal phase is relatively small.

Comparing **P** distributions in **Figs. 7-9**, it can be seen that the dielectric properties of the shell strongly influence the irregular vortex-like distributions in the rhombohedral phase (below 250 K), which are characterized by all three nonzero components of **P**. The dielectric properties of the shell have practically no effect on the structure of polarization single-vortex in the orthorhombic phase (above 250 K), which is characterized by two nonzero components of **P**. The shell has little effect on the disappearance of polarization dynamics upon transition from tetragonal (above 350 K) to the paraelectric phase (above 390 K). Note that our calculations are performed for a particular ferroelectric material, but there are plenty of ferroelectrics for which a shell may affect the core above 390K.

The nanoparticle core in the polymer shell, which poorly screens the polarization due to the small dielectric permittivity, is characterized by multiple irregular shape vortex-like flux-closure domains in the rhombohedral phase and by a sharp transition from the orthorhombic to the paraelectric phase (see **Fig. 7**). Actually, multiple vortices of three polarization components and flux-closure domains exist in XZ-sections near the nanoparticle surface at 50K. They gradually penetrate inside the core at $T$ = (150 – 200) K. It is seen from XY-sections that the vortices and flux-closure domains are counter-directed, they exist near the surface of the nanoparticle and disappear in the equatorial region. With increasing temperature, the vortices penetrate deeper into the core, and for temperatures above 200 K they are directed parallel to the equator. As the temperature increases, the size of the vortex-like areas decreases, leading to a more uniform distribution of the polarization vector. For temperatures above 250 K, a "single" vortex appears simultaneously with complete disappearance of the component $P_2$. Note the spatial distribution of $P_1$ nontrivially depends on temperature. There are two domains of $P_1$, which are weakly temperature dependent for $T < 250$ K, but, starting from 250 K the spatial distribution of $P_1$ depends more strongly on temperature (see XY-sections). In the temperature range (250 – 350) K the component $P_1$ exists only between the poles. For temperatures above 390 K, the region of nonzero $P_1$ expands in the equatorial part and shifts further from the poles of the nanoparticle. The behavior of $P_3$ spatial distribution looks opposite to $P_1$. As the temperature rises above 250K, the component $P_3$ is absent between the poles, and starting from 350 K it gradually vanishes in the equatorial region, tending to disappear throughout the entire the core. Hence, starting from 350 K, the region of zero $P_3$ increases near the poles, while the region of zero $P_1$ increases in the equatorial segment with increasing temperature. For temperatures above 390 K the polarization becomes small and gradually vanishes pointing on the diffuse transition to the paraelectric phase.

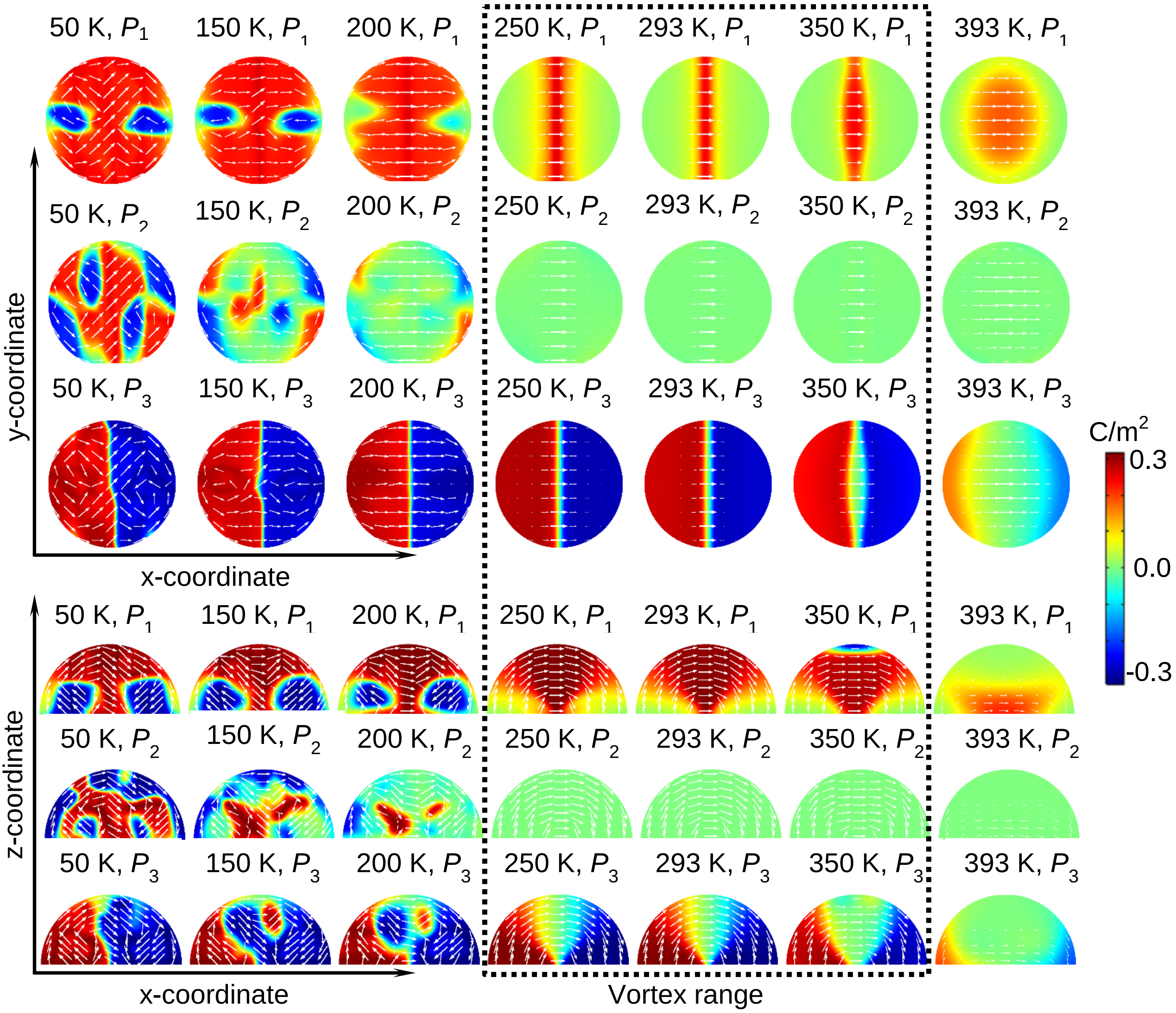

**FIGURE 7.** Polarization components, $P_i(x,y,z)$, inside a $BaTiO_3$ nanoparticle covered with a polymer shell having isotropic temperature-independent dielectric permittivity $\varepsilon_P^S = 3$. The equatorial XY-section (at $z$ = 0) and polar XZ-section (at $y$ = 0) are shown in the plots. Different plots correspond to the temperatures $T$= 50, 150, 200, 250, 293, 350, and 390 K indicated above the graphs. The polarization vector direction is shown by white arrows. The pronounced single-vortex appeared in the temperature range inside the dotted rectangle. $BaTiO_3$ parameters are listed in **Table AI, Appendix A.** Particle radius $R$ = 10 nm, shell thickness $\Delta R$ = 5 nm, and its effective screening length $R_d = 2$ nm.

For a nanoparticle in a paraelectric shell that effectively screens the core polarization due to the high shell's dielectric permittivity, various effects occur in different phases. Namely, the "regular-shape" closure bi-domains with smooth thin walls are inherent to the rhombohedral phase; a single-vortex is observed in orthorhombic phase, and the polarization vortex gradually disappears under the transition to the paraelectric phase (**Fig. 8**). Actually, it can be seen from XY-sections at 50 K that bi-domains are directed along the equator and diverge (or converge) towards the poles for the left (or right) half of the particle, respectively. This behavior is due to the presence of $P_2$ between the poles. Starting from 150 K the vortex polarization becomes parallel to the equator, and simultaneously the component $P_2$ vanishes in the particle. For temperatures (50 – 250) K the magnitude of the component $P_1$ is almost unchanged, and its spatial distribution becomes veryinhomogeneous at temperatures above 250 K. The component $P_1$ is absent in the relatively thin outer equatorial segment of the particle and exists in a wide region between the poles at 250 K. The area of nonzero $P_1$ between the poles narrows with increasing temperature to 350 K, however it expands again at 390 K and fills the entire core under the transition to the paraelectric phase. Despite the oscillating behavior of $P_1$, it does not change its sign. The temperature dependence of the $P_3$ distribution looks opposite to $P_1$. A single region with $P_3$=0 is located between the poles below 350 K, and expands near the equator above 350 K. It can be seen from the XZ-section that the region of zero $P_3$ increases near the pole with increasing temperature above 250 K. At the same time, the region of zero $P_1$ increases near the equator at T ≥ 250 K. Eventually, the polarization remains only in the equatorial segment at T ≥ 390 K, and completely vanishes at $T$ > 400 K corresponding to the paraelectric phase.

For a nanoparticle in the anisotropic LC shell that effectively screens the polarization only in one direction $z$ (coinciding with the LC director orientation), "irregular" multi-domain states are formed in the rhombohedral phase (**Fig. 9**). From the XY-section at $T$ = 200 K, it is seen that the domains are directed along the equator. This behavior is due to the absence of $P_2$ between the poles. For temperatures below 250 K, the component $P_1$ is nearly constant, but its spatial distribution becomes strongly inhomogeneous for temperatures above 250 K. For $T$ = 250 K, the $P_1$ component is absent near the outer equatorial segment of the core, but exists in a wide region between the poles. With a temperature increase to 350 K the area between the poles narrows, however, it apparently expands again at $T$=390 K. Note that despite the oscillating behavior of the component $P_1$, it does not change the sign. The temperature dependence of the $P_3$ spatial distribution looks opposite to $P_1$. For temperatures below 350 K the region of zero $P_3$ is located between the poles, but expands near the equator for $T$ > 350 K.

As seen from the XZ-section at $T$ < 250 K (shown in the bottom part of **Fig. 9**), the distortion of the vortex structure is due to the presence of domains inside the core. At temperatures $T$ > 250K the area of the region with $P_3$ =0 increases near the poles with increasing temperature. At the same time, the region of zero $P_1$ increases near equator with $T$ increase. The polarized region shrinks to the equatorial plane and eventually vanishes for $T$ > 400 K indicating the transition to the paraelectric phase.

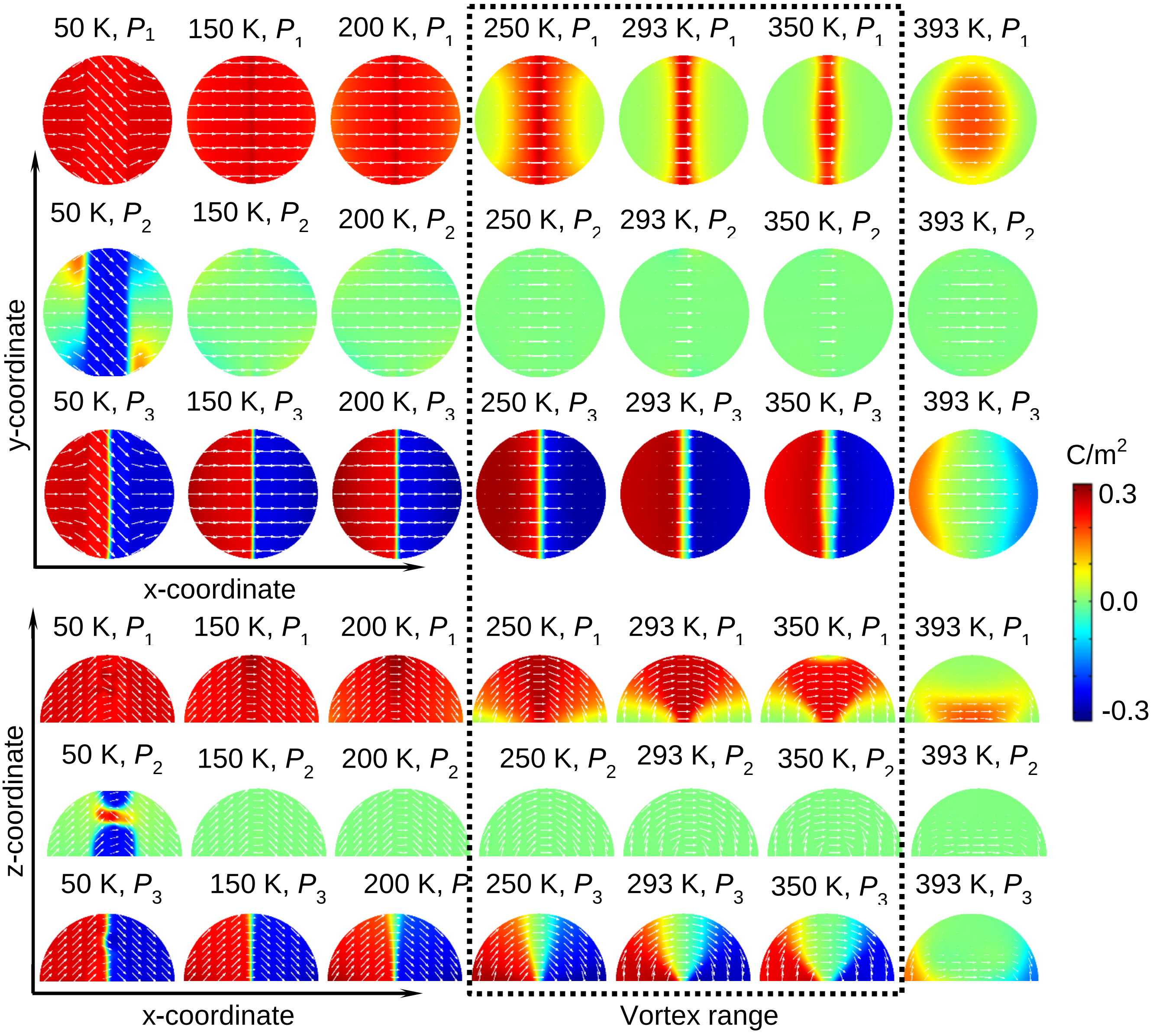

**FIGURE 8.** Polarization components, $P_i(x,y,z)$, inside a $BaTiO_3$ nanoparticle covered with a paraelectric $SrTiO_3$ shell with isotropic temperature-dependent dielectric permittivity given by Eq.(7b). The equatorial XY-section (at $z$ = 0) and polar XZ-section (at $y$ = 0) are shown in the plots. Different plots correspond to the temperatures $T$=50, 150, 200, 250, 293, 350, and 390 K indicated above the graphs. The polarization vector direction is shown by white arrows. The pronounced polarization vortex appeared in the temperature range inside the dotted rectangle. $BaTiO_3$ parameters are listed in **Table AI, Appendix A.** Parameters $R$, $\Delta R$, and $R_d$ are the same as in **Fig. 7.**

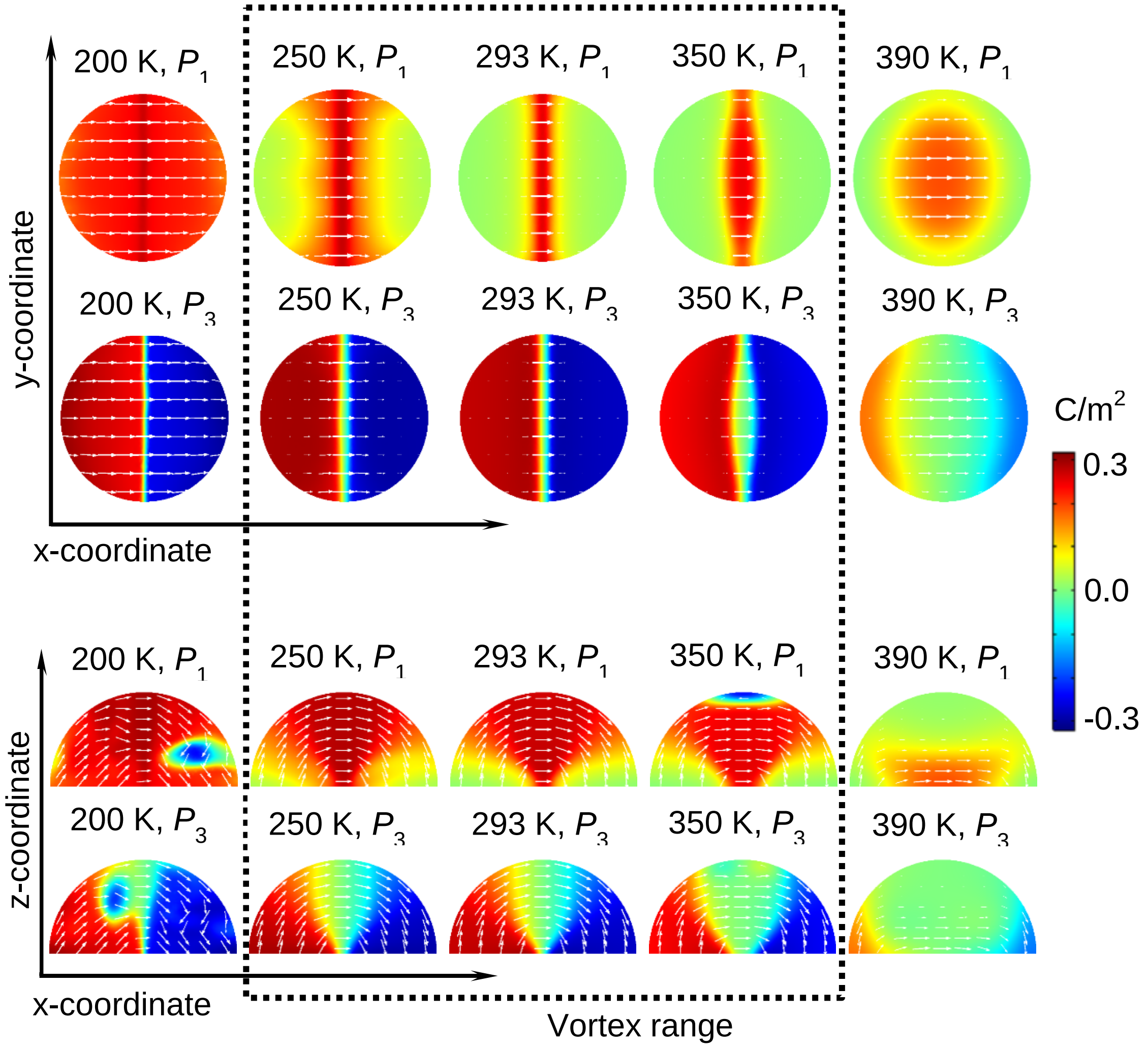

**FIGURE 9.** Polarization components, $P_i(x,y,z)$, inside a $BaTiO_3$ nanoparticle covered with a LC shell having an anisotropic temperature-dependent dielectric permittivity given by Eq.(7c). The equatorial XY-section (at $z = 0$) and the polar XZ-section (at $y = 0$) are shown in the plots. Different plots correspond to the temperatures $T$= 200, 250, 293, 350, and 390 K indicated above the graphs. The polarization vector direction is shown by white arrows. The shell anisotropy axis is parallel to the particle crystallographic axis 3. The pronounced polarization vortex appeared in the temperature range inside the dotted rectangle. $BaTiO_3$ parameters are listed in **Table AI, Appendix A.** Parameters $R$, $\Delta R$, and $R_d$ are the same as in **Fig. 7.**

It turns out that the vortex distribution of the ferroelectric polarization is extremely stable in a wide temperature range 250 K < $T$ < 350 K [indicated by the dotted rectangle in **Figs. 7-9**]. The axis of the vortex rotation is determined by the initial conditions, and can equiprobably coincide with any of crystallographic axes 1, 2 or 3. We varied the screening length $R_d$ of the shell over a wide range from 0.1 nm up to 10 nm, which practically had no affect on the formation of the polarization vortex in the orthorhombic phase. The polarization of the nanoparticle approached the single-domain state only at $R_d$ < 0.01 nm, which corresponds to an almost ideal electrode deposited on the surface of the particle, not the considered polymer, paraelectric, or LC shells with $R_d > 1$ nm and dielectric permittivity described by Eqs. (7). The reason for the vortex stability is that the electric depolarization field is concentrated inside the core, without penetrating the shell. This conclusion will be supported by the following analysis of electrostatic potential φ distribution for different types of shells.

We also shall conclude that the vortex parameters are weakly dependent on the dielectric permittivity of the shell and its screening length. The conclusion will become evident from the analysis of the vorticity degree – toroidal moment $\vec{M} = [\vec{P} \times \vec{r}]$ [74]. Since $P_2$=0 in the vortex-like orthorhombic phase, polarization vortexes can be characterized by the distribution of the component $M_y = zP_1 - xP_3$ and the region of its existence can be compared for different shells.

Distributions of the electrostatic potential φ and normalized toroidal moment component $M_y$ in the polar ZX-section of the $BaTiO_3$ nanoparticle covered with a dielectrically isotropic polymer, tunable isotropic paraelectric $SrTiO_3$, and anisotropic LC shells are shown in **Fig. 10.** Different plots correspond to several temperatures indicated above the graphs. Comparing the φ and $M_y$ distributions, it can be seen that the dielectric properties of the shell strongly influence these distributions in the rhombohedral phase (approximately below 250 K), have practically no effect on the structure of the vortex in the orthorhombic phase (from 250 to 350 K), and have little effect on the dynamics of the polarization disappearance upon the transition from tetragonal (above 350 K) to paraelectric phase (above 390 K). Moreover, the polarization of the particle in a polymer shell, which poorly screens the polarization due to the small dielectric permittivity, is characterized by "double" closing vortices in the rhombohedral phase, as well as a the sharp transition from the orthorhombic to paraelectric phase. For a particle in a paraelectric shell, which screens polarization well due to the large dielectric permittivity, the "regular-shape" closure bi-domains with straight thin walls are inherent to the rhombohedral phase; and the smooth disappearance of the polarization vortex at the transition to the paraelectric phase are pronounced. For a particle in the anisotropic shell, which screens the polarization well only in one crystallographic direction (z-direction) and has a large dielectric permittivity, "irregular" multi-domain states are formed in the rhombohedral phase.

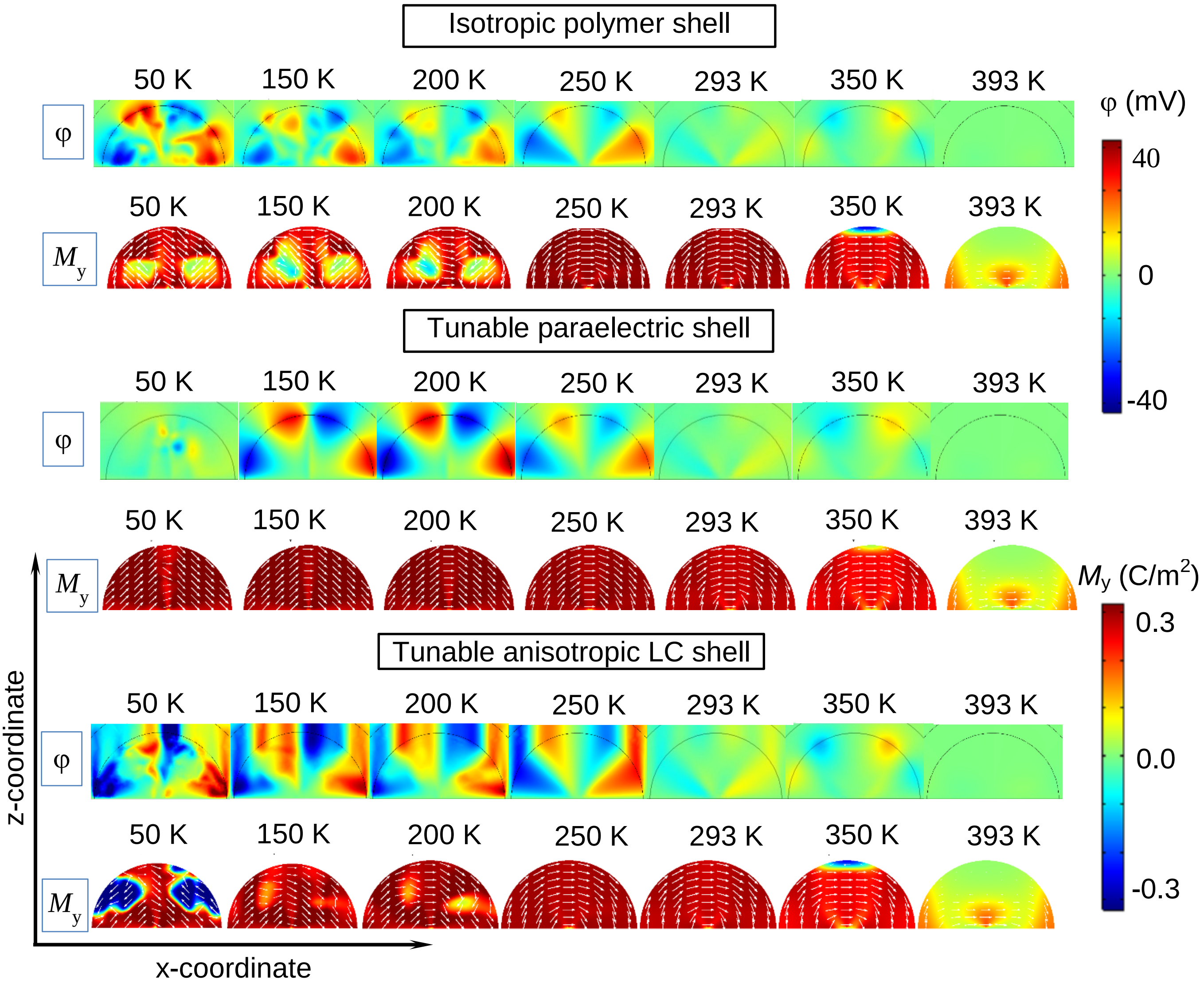

**FIGURE 10.** Distributions of electrostatic potential φ and normalized toroidal moment component $M_y$ in the ZX-section (at $y$ = 0) of $Sn_2P_2S_6$ nanoparticle covered with: a dielectrically isotropic polymer shell ($\varepsilon_P^S = 3$) (top two lines), a tunable isotropic paraelectric shell (middle two lines) and an anisotropic LC shell with $\varepsilon_{zz}$= $\varepsilon_{LC}(T)$ (bottom two lines). The shell anisotropy axis 3 is perpendicular to the particle polarization $P_3$. Different plots correspond to the temperatures $T$= 50, 150, 200, 250, 293, 350, and 390 K indicated above the graphs. Parameters R, $\Delta R$, and $R_d$ are the same as in **Fig. 7.**

Temperature dependences of the average toroidal moment $\vec{M} = \frac{1}{V}\int_V \left[\vec{P}\times\vec{r}\right] d^3r$ and average normalized vorticity, introduced as $p_y = \frac{1}{V}\int_V \left[\vec{P}\times\frac{\vec{r}}{r}\right]_y d^3r$, are shown in **Fig. 11.** Different symbols, calculated by FEM, correspond to a polymer, tunable paraelectric, and anisotropic LC shells. Solid curves are a guide to the eye. The dependences allow determining the temperature regions of rhombohedral, vortex-type orthorhombic, and tetragonal ferroelectric phases, which are separated by dashed vertical lines. The similarity between the average $p_y$ and $M_y$ corroborate the fact that the latter is an appropriate characteristic of polarization vorticity degree.

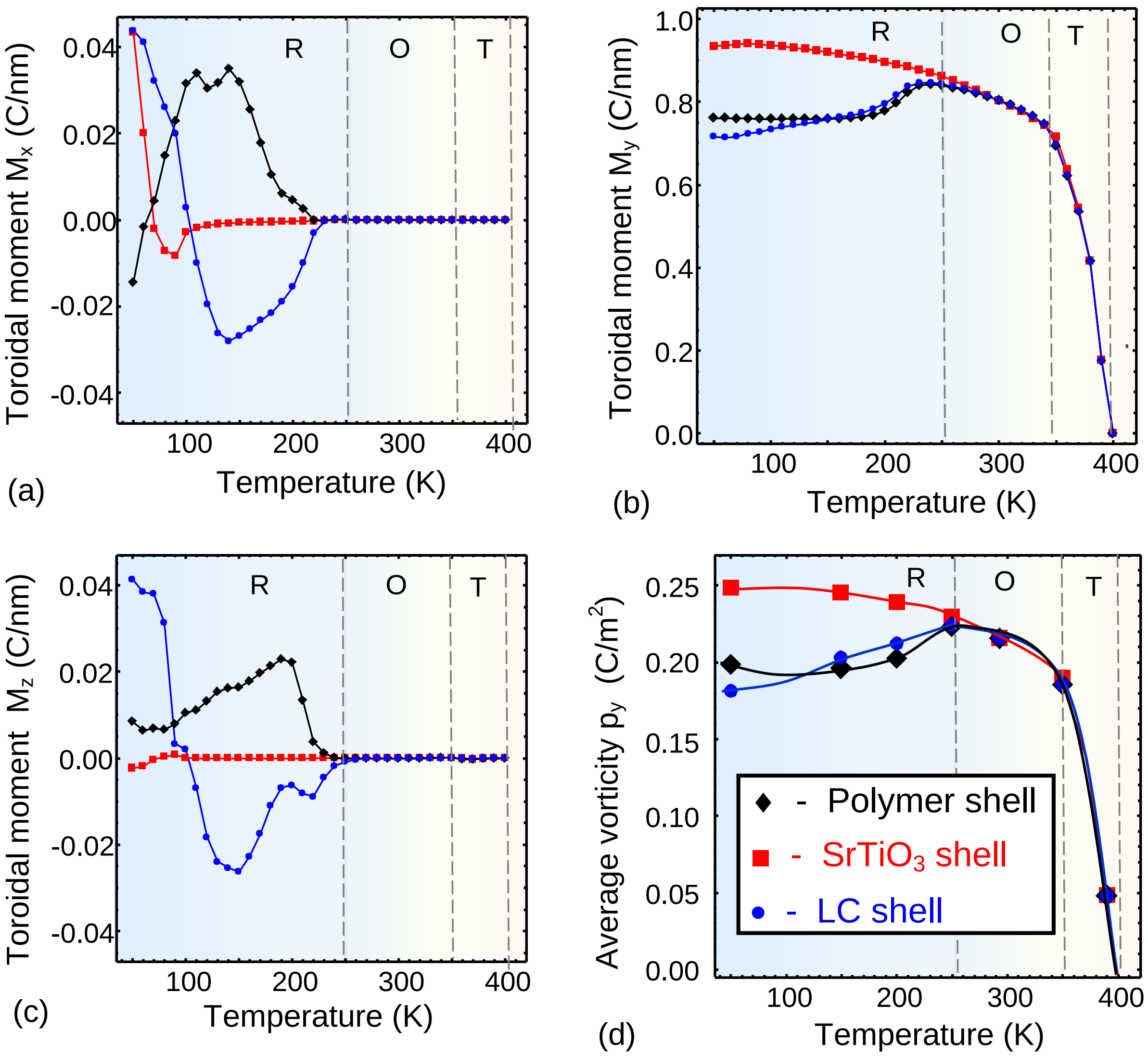

**FIGURE 11.** Temperature dependence of the toroidal moment components, $M_x$ **(a)**, $M_y$ **(b)**, $M_z$ **(c)**, and normalized vorticity $p_y$ **(d)** calculated for a $BaTiO_3$ nanoparticle covered with a dielectrically isotropic polymer (black diamonds), tunable isotropic paraelectric $SrTiO_3$ (red boxes) and anisotropic LC-type (blue circles) shells. The data points denoted by symbols are calculated by FEM. Dashed vertical lines separate the temperature regions of rhombohedral (**R**), orthorhombic (**O**) and tetragonal (**T**) ferroelectric phases. The PE phase transition occurs above 400 K. Parameters $R$, $\Delta R$, and $R_d$ are the same as in **Fig. 7.**

It is seen from **Figs. 11** that all three components of the nanoparticle toroidal polarization moment are non-zero below 250 K, and their temperature dependence strongly depends on the shell material. For a particle in a polymer shell, negative $M_x$ first changes its sign, then passes through a positive double maximum at 150 K, followed by a decrease in value with increasing temperature and a subsequent disappearance above 250 K. Positive $M_y$ decreases slightly, and then slightly increases with increasing temperature. Positive $M_z$ reaches a maximum at about 200 K, and then drops sharply to zero with increasing temperature and disappears above 250 K. For a core in a paraelectric shell, a relatively large positive $M_x$ drops sharply to a small negative value at 80 K, then gradually approaches zero with increasing temperature and finally disappears above 250 K, $M_y$ monotonously decreases with increasing temperature, and a very small negative $M_z$ tends to zero as the temperature increases and disappears above 250 K. For a core in an anisotropic shell both components, $M_x$ and $M_z$, starting from high positive values, first decrease sharply and change sign, then reach a negative minimum around 150 K, followed by a non-monotonical trend to zero as the temperature rises and finally disappear above 250 K. The positive *My* value slowly increases monotonically with increasing temperature, but drops sharply at higher temperatures – similar to the polymer shell. Such temperature behavior of **M** below 250 K corresponds to the "mixed" rhombohedral phase, in which polarization vortices and polydomain states can simultaneously coexist in the particle.

At temperatures over the range (250 – 350) K, $M_x = M_z = 0$, and $M_y$ decreases with increasing temperature and is independent of the shell properties. This behavior of **M** corresponds to the single polarization vortex in the orthorhombic phase.

At temperatures above 350 K, $M_x = M_z = 0$, and $M_y$ sharply decreases with increasing temperature, turning to zero at 400 K corresponding to the PE phase transition. The temperature behavior of $M_y$ does not depend on the shell properties. This behavior of **M** is characteristic for the flux-closure domains, which look like the "vertices" of polarization near the poles of the nanoparticle, and are observed in the tetragonal phase. A transition to the PE phase occurs at 400 K.

The next idea was to check whether the vortex axis can be sensitive to the shell anisotropy. Actually, with a decrease in temperature, when the shell becomes strongly anisotropic, we expected a rotation of the vortex so that its axis would coincide with the axis of the shell with the minimal or maximal dielectric constant. Such a rotation (either the polarization of the nanoparticle core rotates or the nanoparticle itself rotates in space with respect to the LC shell) should minimize the electrostatic energy of the depolarization field of the system. However, we did not observe any polarization rotation, and the difference in the free energy of the original system and the system with the rotated polarization vortex turned out to be very small, in the fourth decimal point.

However one can influence the vortex-like domain structure in a ferroelectric nanoparticle by changing the ambient temperature, because the vortex reacts on the temperature changes of the nanoparticle. At high temperatures close to paraelectric phase transition, $T \approx T_{cr}(R)$, the "round" vortex shape becomes more "square" and shrinks, as shown in **Figs. 7–9.** Then the transition to the PE phase occurs with further temperature increase, where the transition temperature $T_{cr}(R)$ depends on the particle size (size-induced phase transition).

We have found that the distribution of the polarization gradient and amplitude inside the vortex strongly depends on the polarization gradient coefficients $g_{ij}$, which we changed by an order of magnitude in comparison with those given in **Table AII**. When $g_{ij}$ decreases by an order of magnitude, the vortex-like polarization rather resembles a flux-closure bi-domain with relatively thin walls, inside of which the $P_1$ component is non-zero and relatively large, approaching a bulk value of 0.27 C/m$^2$ at room temperature. The bi-domain, which is practically insensitive to the dielectric properties of the shell, exists up to 400 K. With an increase of $g_{ij}$ by an order of magnitude the polarization vortex becomes almost spherical with a developed gradient (i.e. "diffuse") and without any hint of a domain walls. The amplitudes of all polarization components become relatively small in comparison with the bulk value with an increasing temperature over the range (320 – 345) K, depending on the particle size. The blurred spherical vortex becomes sensitive to the dielectric properties of the shell near the transition temperature to the paraelectric phase (on the order of 5 K in its vicinity), and the transition itself looks diffuse, which is impossible for bulk barium titanate. Although according to our calculations the possibility of the vortex control by $g_{ij}$ exists, it probably has very limited practical interest since it is hardly possible to change $g_{ij}$ for a specific ferroelectric, and in addition, the nanoparticle polarization depends on the dielectric properties of the shell in a very narrow temperature range.

The influence homogeneous and inhomogeneous external electric fields have on vortex-like domains formed inside a spherical nanoparticle is an important aspect to be considered for applications, and will be studied elsewhere. For example, in electrocaloric applications the polarization response to a small external field should be as high as possible, and so the polarization vortex (as an electric toroidal multipole) seems much less favorable than the single- or poly-domain states (electric dipoles) considered in the previous subsection. To change the toroidal moment of the vortex polarization the curled electric field, $\vec{E}_{cur} = \frac{1}{2}\vec{S}\times\vec{r}$, originated from a quasi-static magnetic field, $rot\vec{E}_{cur} = -\frac{\partial}{\partial t}\vec{B}$, can be applied to the nanoparticle [72, 73, 74, 75, 76]. The “coercive” vorticity vector $S \geq 10^{16}$ V/m$^2$ [74]; therefore, the “coercive” curled electric field should be very high ($\vec{E}_{cur} \geq 10^8$ V/m) at the surface of 10 nm nanoparticle.

## IV. DISCUSSION AND CONCLUSIVE REMARKS

The possibility of controlling the domain structure in spherical nanoparticles with a uniaxial or multiaxial ferroelectric core using a shell with tunable dielectric properties was studied in the framework of LGD theory. FEM was performed for $Sn_2P_2S_6$ and $BaTiO_3$ nanoparticles covered with either temperature-independent high-k polymer, temperature-dependent isotropic paraelectric strontium titanate, or temperature dependent anisotropic LC shell.

It appeared that the "tunable" paraelectric shell with temperature dependent high dielectric permittivity (~300 – 3000) provides much more efficient screening of the nanoparticle polarization in comparison with the polymer shell that has a much smaller (~10) temperature-independent permittivity. The tunable temperature-dependent dielectric anisotropy of the LC shell (~ 1 – 100) adds a new level of functionality for the control of ferroelectric domain structure in comparison with isotropic paraelectric and polymer shells. The conclusion is quantified by analytical calculations of nanoparticle phase diagrams for the case of the $Sn_2P_2S_6$ core covered with considered three types of shells.

It was shown that by varying the magnitude and anisotropy of the shell, one can control the appearance of polarization vortices in $BaTiO_3$ nanoparticle. Namely, if the screening conditions deteriorate (e.g. for an isotropic shell with a small dielectric constant and a large screening length), it is possible to achieve a complete disappearance of both vortices and polar state of the nanoparticle. On the other hand, a half-metallic shell with an ultra-small screening length and very high dielectric constant provides the absolute stability of the single-domain ferroelectric state of the nanoparticle.

Also we established that domain morphology in $Sn_2P_2S_6$ and $BaTiO_3$ nanoparticles behaves principally different, because vortex polarization of the $BaTiO_3$ core is extremely stable to ambient screening conditions, while uniaxial polar state of the $Sn_2P_2S_6$ core is easy to destroy by the tiny changes of the screening conditions. The conclusion is quantified by numerical calculations of the polarization vorticity in a rhombohedral, orthorhombic, and tetragonal phases of a $BaTiO_3$ core for the considered three types of shells.

Since there are real possibilities to change the value and anisotropy of the shell dielectric permittivity, obtained results indicate the opportunities to control the domain structure of uniaxial and multiaxial ferroelectric nanoparticles covered with tunable shells. This may allow for a new generation of ferroelectric memory and advanced cryptographic materials.

**Acknowledgements.** A.N.M. expresses her deepest gratitude to Prof. S.M. Ryabchenko (NASU), and Prof. R. Hertel (CNRS) for stimulating discussions about the nature of effective screening length, electrostatic problems and polarization vortexes.

**Authors' contribution.** A.N.M. and V.Yu.R. generated the research idea, stated the problem, performed analytical calculations, interpreted results and wrote the manuscript draft. E.A.E. jointly with Y.M.F. wrote the codes and performed numerical calculations. Yu.M.V. and D.R.E. densely worked on the results interpretation and manuscript improvement.

## APPENDIX A. Euler-Lagrange equations and material parameters

Allowing for the Khalatnikov mechanism of polarization relaxation, minimization of the free energy (2) with respect to polarization leads to three coupled time-dependent LGD-equations for polarization components, $\frac{\delta G}{\delta P_i}=-\Gamma\frac{\partial P_i}{\partial t}$, where the explicit form for a ferroelectric nanoparticle with m3m parent symmetry is:

$$\begin{aligned}&\Gamma\frac{\partial P_1}{\partial t}+2P_1\left(a_1-Q_{12}\left(\sigma_{22}+\sigma_{33}\right)-Q_{11}\sigma_{11}\right)-Q_{44}\left(\sigma_{12}P_2+\sigma_{13}P_3\right)+4a_{11}P_1^3\\&+2a_{12}P_1\left(P_2^2+P_3^2\right)+6a_{111}P_1^5+2a_{112}P_1\left(P_2^4+2P_1^2P_2^2+P_3^4+2P_1^2P_3^2\right)+2a_{112}P_1P_2^2P_3^2\\&-g_{11}\frac{\partial^2P_1}{\partial x_1^2}-g_{44}\left(\frac{\partial^2P_1}{\partial x_2^2}+\frac{\partial^2P_1}{\partial x_3^2}\right)-\left(g'_{44}+g_{12}\right)\frac{\partial^2P_2}{\partial x_2\partial x_1}-\left(g'_{44}+g_{12}\right)\frac{\partial^2P_3}{\partial x_3\partial x_1}\\&+F_{11}\frac{\partial\sigma_{11}}{\partial x_1}+F_{12}\left(\frac{\partial\sigma_{22}}{\partial x_1}+\frac{\partial\sigma_{33}}{\partial x_1}\right)+F_{44}\left(\frac{\partial\sigma_{12}}{\partial x_2}+\frac{\partial\sigma_{13}}{\partial x_3}\right)=E_1\end{aligned}\tag{A.1a}$$

$$\begin{aligned}&\Gamma\frac{\partial P_2}{\partial t}+2P_2\left(a_1-Q_{12}\left(\sigma_{11}+\sigma_{33}\right)-Q_{11}\sigma_{22}\right)-Q_{44}\left(\sigma_{12}P_1+\sigma_{23}P_3\right)+4a_{11}P_2^3\\&+2a_{12}P_2\left(P_1^2+P_3^2\right)+6a_{111}P_2^5+2a_{112}P_2\left(P_1^4+2P_2^2P_1^2+P_3^4+2P_2^2P_3^2\right)+2a_{112}P_2P_1^2P_3^2\\&-g_{11}\frac{\partial^2P_2}{\partial x_2^2}-g_{44}\left(\frac{\partial^2P_2}{\partial x_1^2}+\frac{\partial^2P_2}{\partial x_3^2}\right)-\left(g'_{44}+g_{12}\right)\frac{\partial^2P_1}{\partial x_2\partial x_1}-\left(g'_{44}+g_{12}\right)\frac{\partial^2P_3}{\partial x_3\partial x_2}\\&+F_{11}\frac{\partial\sigma_{22}}{\partial x_2}+F_{12}\left(\frac{\partial\sigma_{11}}{\partial x_2}+\frac{\partial\sigma_{33}}{\partial x_2}\right)+F_{44}\left(\frac{\partial\sigma_{12}}{\partial x_1}+\frac{\partial\sigma_{23}}{\partial x_3}\right)=E_2\end{aligned}\tag{A.1b}$$

$$\begin{aligned}&\Gamma\frac{\partial P_3}{\partial t}+2P_3\left(a_1-Q_{12}\left(\sigma_{11}+\sigma_{22}\right)-Q_{11}\sigma_{33}\right)-Q_{44}\left(\sigma_{13}P_1+\sigma_{23}P_2\right)+4a_{11}P_3^3\\&+2a_{12}P_3\left(P_1^2+P_2^2\right)+6a_{111}P_3^5+2a_{112}P_3\left(P_1^4+2P_3^2P_1^2+P_2^4+2P_2^2P_3^2\right)+2a_{112}P_3P_1^2P_2^2\\&-g_{11}\frac{\partial^2P_3}{\partial x_3^2}-g_{44}\left(\frac{\partial^2P_3}{\partial x_1^2}+\frac{\partial^2P_3}{\partial x_2^2}\right)-\left(g'_{44}+g_{12}\right)\frac{\partial^2P_1}{\partial x_3\partial x_1}-\left(g'_{44}+g_{12}\right)\frac{\partial^2P_2}{\partial x_3\partial x_2}\\&+F_{11}\frac{\partial\sigma_{33}}{\partial x_3}+F_{12}\left(\frac{\partial\sigma_{11}}{\partial x_3}+\frac{\partial\sigma_{33}}{\partial x_3}\right)+F_{44}\left(\frac{\partial\sigma_{13}}{\partial x_1}+\frac{\partial\sigma_{23}}{\partial x_2}\right)=E_3\end{aligned}\tag{A.1c}$$

The Khalatnikov coefficient $\Gamma$ determines the relaxation time of polarization $\tau_K=\Gamma/|\alpha|$, which typically varies in the range ($10^{-11}$ – $10^{-13}$)s for temperatures far from $T_C$. The boundary condition for polarization at the core-shell interface $r=R$ is natural, but accounts for the flexoelectric effect:

$$a^{(P)}P_i+\left(g_{ijkl}\frac{\partial P_k}{\partial x_l}-F_{klij}\sigma_{kl}\right)n_j\Bigg|_{r=R}=0\qquad(i=1,2,3)\tag{A.2}$$

where **n** is the outer normal to the surface.

Elastic stresses satisfy the equation of mechanical equilibrium in the nanoparticle and its shell,

$$\frac{\partial \sigma_{ij}}{\partial x_j} = 0 , \; 0 < r < R + \Delta R . \qquad \text{(A.3a)}$$

Equations of state should be obtained from the variation of the energy (2e) with respect to elastic stress, $\frac{\delta G_{es+flexo}}{\delta \sigma_{ij}} = -u_{ij}$, namely:

$$s_{ijkl}\sigma_{ij} + Q_{ijkl}P_k P_l + F_{ijkl}\frac{\partial P_l}{\partial x_k} = -u_{ij} , \quad 0 < r < R \qquad \text{(A.3b)}$$

$$s^S_{ijkl}\sigma_{ij} = -u_{ij} , \quad R < r < R + \Delta R \qquad \text{(A.3b)}$$

where $u_{ij}$ is the strain tensor, $u_{ij} = \left(\partial U_i / \partial x_j + \partial U_j / \partial x_i\right)/2$, and the displacement vector components are $U_i$. In principle we can assume that $s^S_{ijkl} \approx s_{ijkl}$ for a "soft" shell.

Elastic boundary conditions at the particle core-shell interface $r = R + \Delta R$ are the continuity of the normal elastic stresses:

$$\sigma_{rr}\big|_{r=R-0} = \sigma_{rr}\big|_{r=R+0} , \quad \sigma_{r\theta}\big|_{r=R-0} = \sigma_{r\theta}\big|_{r=R+0} , \quad \sigma_{r\phi}\big|_{r=R-0} = \sigma_{r\phi}\big|_{r=R+0} , \qquad \text{(A.3c)}$$

and displacement vector components

$$U_r\big|_{r=R-0} = U_r\big|_{r=R+0} , \quad U_\theta\big|_{r=R-0} = U_\theta\big|_{r=R+0} , \quad U_\phi\big|_{r=R-0} = U_\phi\big|_{r=R+0} . \qquad \text{(A.3d)}$$

**Table AI.** LGD parameters for bulk ferroelectric $Sn_2P_2S_6$

| $\varepsilon_b$ | $\alpha_T$($C^{-2}$·m J/K) | $T_C$ (K) | β ($C^{-4}$·$m^5$J) | γ ($C^{-6}$·$m^9$J) | $g_{11}$ ($m^3$/F) | $g_{44}$ ($m^3$/F) |
|---|---|---|---|---|---|---|
| 7 | $1.6\times10^6$ | 336 | $7.42\times10^8$ | $3.5\times10^{10}$ | $3.0\times10^{-10}$ | $0.3\times10^{-10}$ |

**Table AII.** Material parameters for bulk ferroelectric $BaTiO_3$

| **coefficient** | **$BaTiO_3$** (collected and recalculated mainly from Ref. [a, b]) |
|---|---|
| **Symmetry** | Tetragonal at room temperature, m3m in a paraelectric phase |
| $\varepsilon_b$ | 7 (Ref. [b]) |
| $a_i$ (**$C^{-2}$·mJ**) | $a_1$=3.34($T$–381)×$10^5$, (at 293°K –2.94×$10^7$) |
| $a_{ij}$ (**$C^{-4}$·$m^5$J**) | $a_{11}$= 4.69($T$–393)×$10^6$–2.02×$10^8$, $a_{12}$= 3.230×$10^8$, (at 293°K $a_{11}$= –6.71×$10^8$ $a_{12}$= 3.23×$10^8$) |
| $a_{ijk}$ (**$C^{-6}$·$m^9$J**) | $a_{111}$= –5.52($T$–393)×$10^7$+2.76×$10^9$, $a_{112}$=4.47×$10^9$, $a_{123}$=4.91×$10^9$ (at 293°K $a_{111}$= 82.8×$10^8$, $a_{112}$=44.7×$10^8$, $a_{123}$=49.1×$10^8$) |
| $Q_{ij}$ (**$C^{-2}$·$m^4$**) | $Q_{11}$=0.11, $Q_{12}$= –0.043, $Q_{44}$=0.059 |
| $s_{ij}$ (**×$10^{-12}$ $Pa^{-1}$**) | $s_{11}$=8.3, $s_{12}$= –2.7, $s_{44}$=9.24 |
| $g_{ij}$ (**×$10^{-10}$$C^{-2}$$m^3$J**) | $g_{11}$=5.1, $g_{12}$= –0.2, $g_{44}$= 0.2 [c] |
| ***F*$_{ij}$ (×10$^{-11}$C$^{-1}$m$^3$)** | ~100 (estimated from measurements of Ref. [d]), $F_{11}$= +2.46, $F_{12}$=0.48, $F_{44}$=0.05 (recalculated from [e] using $F_{\alpha\gamma}=f_{\alpha\beta}s_{\beta\gamma}$) |
| ***R*$_d$ (nm)** | 0.1 |

## APPENDIX B. Derivation of the transition temperature in a single-domain approximation

Let us consider the spherical ferroelectric particle with polarization $\mathbf{P}$ oriented along one of the principal crystallographic axes, z. Here we also introduce an isotropic background permittivity $\varepsilon_b$ of the ferroelectric particle. The media outside the particle is a dielectric with permittivity $\varepsilon_e$. Electrical displacement is $\mathbf{D}_i=\varepsilon_0\varepsilon_b\mathbf{E}+P\mathbf{e}_z$, $\mathbf{D}_S=\varepsilon_0\hat{\varepsilon}_S\mathbf{E}$ and $\mathbf{D}_e=\varepsilon_0\varepsilon_e\mathbf{E}$, where the subscript "*i*" means the physical quantity inside the particle, "s" – inside the shell and "*e*" – outside the particle; $\varepsilon_0$ is a universal dielectric constant. Hereinafter we suppose that the polarization gradient is small.

Below we consider either dielectrically isotropic shell ($\varepsilon_{ij}^S=\varepsilon^S\delta_{ij}$) or LC shell with a special type of "hedgehog-like" anisotropy with nonzero components $\varepsilon_{rr}=\varepsilon_{\parallel}$ and $\varepsilon_{\theta\theta}=\varepsilon_{\phi\phi}=\varepsilon_{\perp}$. The anisotropy corresponds to the dominant role of Van-der-Waals interactions between banana-shaped LC molecules and the ferroelectric surface. The polarization effect is regarded as the next-order correction.

We introduce an electric field $\mathbf{E}=-\nabla\varphi$ via electrostatic potential $\varphi$ that should satisfy the Poisson equation inside the particle, Debye-type equation inside the shell, and Laplace equation outside it, respectively:

$$\varepsilon_0\varepsilon_b\Delta\varphi_i=-\frac{\partial P}{\partial z},\qquad 0\le r<R, \tag{B.1a}$$

$$\Delta_S\varphi_S-\frac{\varphi_S}{R_d^2}=0,\qquad R<r<R+\Delta R, \tag{B.1b}$$

$$\Delta\varphi_e=0,\qquad r>R+\Delta R. \tag{B.1c}$$

For a spherical particle with an isotropic dielectric, the Laplace operator in Eq.(B.1a,c) in spherical coordinates has the usual form, $\Delta=\frac{1}{r^2}\frac{\partial}{\partial r}\left(r^2\frac{\partial}{\partial r}\right)+\frac{1}{r\sin\theta}\frac{\partial}{\partial\theta}\left(\sin\theta\frac{1}{r}\frac{\partial}{\partial\theta}\right)+\frac{1}{r^2\sin^2\theta}\frac{\partial^2}{\partial\phi^2}$. For a shell that can be dielectrically anisotropic, the "effective" Laplace operator was introduced in Eq.(B.1b) as $\Delta_S = \frac{1}{r^2}\frac{\partial}{\partial r}\left(r^2\frac{\partial}{\partial r}\right) + \frac{1}{r\sin\theta}\frac{\partial}{\partial\theta}\left(\sin\theta\frac{\varepsilon_\perp}{\varepsilon_\parallel}\frac{1}{r}\frac{\partial}{\partial\theta}\right) + \frac{1}{r^2\sin^2\theta}\frac{\varepsilon_\perp}{\varepsilon_\parallel}\frac{\partial^2}{\partial\phi^2}$. Here $\theta$ is the polar angle, $\phi$ is azimuthal angle, $r$ is the radial coordinate.

Equations (B.1) should be supplemented by the interface conditions for potential continuity at all surfaces for electric potential and normal components of displacement:

$$\varphi_i\big|_{r\to 0} < \infty, \quad (\varphi_s - \varphi_i)\big|_{r=R} = 0; \quad (\varphi_e - \varphi_S)\big|_{r=R+\Delta R} = 0, \quad \varphi_e\big|_{r\to\infty} = -z\,E_{ext} \equiv -r\cos\theta E_{ext} \tag{B.2a}$$

$$(\mathbf{D}_S - \mathbf{D}_i)\mathbf{e}_r \equiv (D_{Sr} - \varepsilon_0\varepsilon_b E_{ri} - P\cos\theta)\big|_{r=R} = 0, \ (\mathbf{D}_e - \mathbf{D}_S)\mathbf{e}_r \equiv (D_{er} - \varepsilon_0\varepsilon_e E_{er})\big|_{r=R+\Delta R} = 0. \tag{B.2b}$$

Here $\mathbf{e_r}$ is the outer normal to the particle surface. In Eq. (B.2a) $E_{ext}$ is the external electric field far from the particle (if any is applied).

For a spherical particle the general solutions of Eq.(B.1) could be expanded into the series of the Legendre polynomials, and the azimuthal-angle independent solution has the form $\varphi(r,\theta) \sim f(r)\cos\theta$ in each region, where the radial function $f(r)$ obeys the equations:

$$r^2\frac{\partial^2 f}{\partial r^2} + 2r\frac{\partial f}{\partial r} - 2f = 0, \qquad 0 \le r < R \quad \text{and} \quad r > R + \Delta R, \tag{B.3a}$$

$$r^2\frac{\partial^2 f}{\partial r^2} + 2r\frac{\partial f}{\partial r} - \frac{2\varepsilon_\perp}{\varepsilon_\parallel} f - \frac{r^2}{R_d^2} f = 0, \qquad R < r < R + \Delta R. \tag{B.3b}$$

Solution to Eqn.(B.3a) can be easily found, $f = Ar$ inside the particle and $f = \frac{B}{r^2} + Cr$ outside the shell:

$$\varphi_i = -E_i r\cos\theta, \qquad 0 \le r < R, \tag{B.4a}$$

$$\varphi_e = \left(E_e\frac{R^3}{r^2} - r\,E_{ext}\right)\cos\theta, \qquad r > R + \Delta R. \tag{B.4b}$$

The solution to the Eq.(B.3b) is $f(\rho) = E_S^+\sqrt{\frac{2}{\pi\rho}}I_\zeta(\rho) + E_S^-\sqrt{\frac{2}{\pi\rho}}K_\zeta(\rho)$, where $\rho = \frac{r}{R_d}$, $\zeta = \sqrt{\frac{1}{4} + 2\frac{\varepsilon_\perp}{\varepsilon_\parallel}}$, also here we introduced modified Bessel functions, which can be expressed via ordinary Bessel functions as follows:

$$\varphi_S = \left(E_S^+\sqrt{\frac{2R_d}{\pi r}}I_{\sqrt{\frac{1}{4}+2\frac{\varepsilon_\perp}{\varepsilon_\parallel}}}\left(\frac{r}{R_d}\right) + E_S^-\sqrt{\frac{2R_d}{\pi r}}K_{\sqrt{\frac{1}{4}+2\frac{\varepsilon_\perp}{\varepsilon_\parallel}}}\left(\frac{r}{R_d}\right)\right)R\cos\theta, \qquad R < r < R + \Delta R. \tag{B.4c}$$

The constants $E_i$, $E_S^\pm$, and $E_e$ should be determined from the boundary conditions (B.2).

**B1. Approximate solution for isotropic semiconducting shell in an external field**

In particular case of an isotropic shell, $\varepsilon_\perp = \varepsilon_\parallel = \varepsilon_S$, the solution of Eq.(B.3c) reads $f(\rho) = \frac{E_J}{\rho^2}[\cosh(\rho) - \rho\sinh(\rho)] + \frac{E_Y}{\rho^2}[\rho\cosh(\rho) - \sinh(\rho)]$, where $\rho = \frac{r}{R_d}$, $E_j$ and $E_Y$ are integration constants. Elementary transformations lead to $f(\rho) = \frac{\exp(\rho)}{\rho^2}(1-\rho)E_S^+ + \frac{\exp(-\rho)}{\rho^2}(1+\rho)E_S^-$, so that:

$$\varphi_S = \frac{R_d^2}{r^2}\left(\exp\left(\frac{r}{R_d}\right)\left(1 - \frac{r}{R_d}\right)E_S^+ + \exp\left(-\frac{r}{R_d}\right)\left(1 + \frac{r}{R_d}\right)E_S^-\right)R\cos\theta. \tag{B.5}$$

Four constants $E_i$, $E_S^\pm$, and $E_e$ should be determined from the boundary conditions (B.2). Namely the potential continuity at $r = R$ and $r = R + \Delta R$ gives two equations:

$$E_i = -\frac{R_d^2}{R^2}\left(\exp\left(\frac{R}{R_d}\right)\left(1 - \frac{R}{R_d}\right)E_S^+ + \exp\left(-\frac{R}{R_d}\right)\left(1 + \frac{R}{R_d}\right)E_S^-\right), \tag{B.6a}$$

$$\begin{gathered}\frac{R_d^2 R}{(R+\Delta R)^2}\left(\exp\left(\frac{R+\Delta R}{R_d}\right)\left(1 - \frac{R+\Delta R}{R_d}\right)E_S^+ + \exp\left(-\frac{R+\Delta R}{R_d}\right)\left(1 + \frac{R+\Delta R}{R_d}\right)E_S^-\right) \\ = \frac{R^3 E_e}{(R+\Delta R)^2} - (R+\Delta R)E_{ext}\end{gathered}. \tag{B.6b}$$

Using the expressions for the radial components of electric displacements, $(\mathbf{D}_i)_r = (\varepsilon_0\varepsilon_b E_i + P)\cos\theta$, $(\mathbf{D}_S)_r = \varepsilon_0\varepsilon_S\left[E_S^-\left(2 + \frac{2r}{R_d} + \frac{r^2}{R_d^2}\right)\exp\left(-\frac{r}{R_d}\right) + E_S^+\left(2 - \frac{2r}{R_d} + \frac{r^2}{R_d^2}\right)\exp\left(\frac{r}{R_d}\right)\right]\frac{R^3}{r^3}\cos\theta$, and $(\mathbf{D}_e)_r = \varepsilon_0\varepsilon_e\left[2E_e\frac{R^3}{r^3} + E_{ext}\right]\cos\theta$, from the conditions on the normal components electric displacements continuity, we obtained two more equations:

$$\varepsilon_0\varepsilon_S\left[E_S^-\left(2 + \frac{2R}{R_d} + R\frac{r^2}{R_d^2}\right)\exp\left(-\frac{R}{R_d}\right) + E_S^+\left(2 - \frac{2R}{R_d} + \frac{R^2}{R_d^2}\right)\exp\left(\frac{R}{R_d}\right)\right] = \varepsilon_0\varepsilon_b E_i + P, \tag{B.6c}$$

$$\varepsilon_e\left[\frac{2E_e R^3}{(R+\Delta R)^3} + E_{ext}\right] = \frac{\varepsilon_S R^3}{(R+\Delta R)^3}\left[\begin{gathered}E_S^-\left(2 + \frac{2(R+\Delta R)}{R_d} + \frac{(R+\Delta R)^2}{R_d^2}\right)\exp\left(-\frac{R+\Delta R}{R_d}\right) \\ + E_S^+\left(2 - \frac{2(R+\Delta R)}{R_d} + \frac{(R+\Delta R)^2}{R_d^2}\right)\exp\left(\frac{R+\Delta R}{R_d}\right)\end{gathered}\right] \tag{B.6d}$$

From these equations one could find two contributions to the electric field inside the particle,

$$E_i = E_i^{dep} + E_i^{ext}, \tag{B.7a}$$

where the depolarization field:

$$E_i^{dep} = -\frac{P}{\varepsilon_0}\eta(R, \Delta R, R_d), \tag{B.8a}$$

is defined by a depolarization factor

$$\eta(R,\Delta R,R_d)=\frac{R_d}{Det(R,\Delta R,R_d)}\left\{\begin{matrix}\exp\left(\frac{\Delta R}{R_d}\right)(R_d+R)\left[2R_d(R+\Delta R-R_d)(\varepsilon_e-\varepsilon_S)+(R+\Delta R)^2\varepsilon_S\right]-\\ \exp\left(-\frac{\Delta R}{R_d}\right)(R_d-R)\left[2R_d(R+\Delta R+R_d)(\varepsilon_S-\varepsilon_e)+(R+\Delta R)^2\varepsilon_S\right]\end{matrix}\right\} \quad \text{(B.8b)}$$

$$Det(R,\Delta R,R_d)=\left(\begin{matrix}\exp\left(\frac{\Delta R}{R_d}\right)\left[R_d(R_d+R)(\varepsilon_b+2\varepsilon_S)+R^2\varepsilon_S\right]\left[2R_d(R+\Delta R-R_d)(\varepsilon_e-\varepsilon_S)+(R+\Delta R)^2\varepsilon_S\right]-\\ \exp\left(-\frac{\Delta R}{R_d}\right)\left[R_d(R_d-R)(\varepsilon_b+2\varepsilon_S)+R^2\varepsilon_S\right]\left[2R_d(R+\Delta R+R_d)(\varepsilon_S-\varepsilon_e)+(R+\Delta R)^2\varepsilon_S\right]\end{matrix}\right)$$

(B.8c)

and an external electric field, screened by the shell:

$$E_i^{ext}=\frac{6R_d(R+\Delta R)^3}{Det(R,\Delta R,R_d)}\varepsilon_e\varepsilon_S E_{ext} \quad \text{(B.8d)}$$

In a particular case $|\varepsilon_e-\varepsilon_S|<<\varepsilon_S$ or/and $(R+\Delta R)>>R_d$, Eq.(B.8) simplifies to:

$$E_i^{dep}=\frac{-\frac{P}{\varepsilon_0}R_d\left\{(R_d+R)-\exp\left(-2\frac{\Delta R}{R_d}\right)(R_d-R)\right\}}{\left(R_d(R_d+R)(\varepsilon_b+2\varepsilon_S)+R^2\varepsilon_S-\exp\left(-2\frac{\Delta R}{R_d}\right)\left[R_d(R_d-R)(\varepsilon_b+2\varepsilon_S)+R^2\varepsilon_S\right]\right)} \quad \text{(B.9a)}$$

$$\approx-\frac{P}{\varepsilon_0}\left[\varepsilon_b+2\varepsilon_S+\frac{R^2\varepsilon_S}{R_d(R_d+R)}\right]^{-1}$$

$$E_i^{ext}=\frac{6R_d(R+\Delta R)\exp\left(-\frac{\Delta R}{R_d}\right)\varepsilon_e E_{ext}}{R_d(R_d+R)(\varepsilon_b+2\varepsilon_S)+R^2\varepsilon_S-\exp\left(-2\frac{\Delta R}{R_d}\right)\left[R_d(R_d-R)(\varepsilon_b+2\varepsilon_S)+R^2\varepsilon_S\right]}$$

(B.9b)

$$\approx\frac{6R_d(R+\Delta R)\exp\left(-\frac{\Delta R}{R_d}\right)\varepsilon_e E_{ext}}{R_d(R_d+R)(\varepsilon_b+2\varepsilon_S)+R^2\varepsilon_S}\sim\frac{6\varepsilon_e E_{ext}}{(\varepsilon_b+2\varepsilon_S)R_d+\varepsilon_S R}\exp\left(-\frac{\Delta R}{R_d}\right)$$

Approximate equalities in Eqs.(B.9) are valid at $\Delta R>>R_d$.

Below we use the expression (B.8) for the formulation of the phenomenological equations of state. Substituting Eqs.(B.10) into the LGD equation one obtains the equation for the polarization of the spherical particle

$$\left(\alpha_T(T-T_C^*)+\frac{\eta(R,\Delta R,R_d)}{\varepsilon_0}\right)P+\beta P^3=E_i^{ext}(R,\Delta R,R_d) \quad \text{(B.10)}$$

Approximate expression for the nanoparticle transition temperature $T_{cr}$ from SDFE to PE phase follows from Eq.(B.10):

$$T_{cr}(R)=T_C^*-\frac{\eta(R,\Delta R,R_d)}{\alpha_T\varepsilon_0}, \tag{B.11}$$

### B2. Approximate solution for an isotropic thick semiconducting shell without an external field

In particular case of well-conductive and/or enough thick isotropic shell, $\Delta R>>\frac{R_d}{\varepsilon^S}$ and $\varepsilon_\perp=\varepsilon_\|=\varepsilon_S$, and under the absence of external field, $E_{ext}=0$, Eq.(B.4c) simplifies as

$$\varphi_S=E_S\left(\frac{R^3}{r^2}+\frac{R^3}{R_d r}\right)\exp\left(-\frac{r-R}{R_d}\right)\cos\theta \tag{B.12}$$

Substitution of Eqs.(B.4) to the boundary conditions Eqs. (B.2) leads to condition

$$E_S\frac{R_d+R}{R_d}R\cos\theta=-E_i R\cos\theta \quad\Rightarrow\quad E_i=-E_S\frac{R_d+R}{R_d}. \tag{B.13}$$

Radial components of field could be obtained from (B.3) as follows

$$(\mathbf{E}_S)_r=E_S\frac{R^3}{r^3}\left(2+\frac{2r}{R_d}+\frac{r^2}{R_d^2}\right)\exp\left(-\frac{r-R}{R_d}\right)\cos\theta,\qquad (\mathbf{E}_i)_r=E_i\cos\theta. \tag{B.14}$$

Corresponding displacement is

$$(\mathbf{D}_S)_r=\varepsilon_0\varepsilon_S E_S\frac{R^3}{r^3}\left(2+\frac{2r}{R_d}+\frac{r^2}{R_d^2}\right)\exp\left(-\frac{r-R}{R_d}\right)\cos\theta,\qquad (\mathbf{D}_i)_r=\varepsilon_0\varepsilon_b E_i\cos\theta+P\cos\theta \tag{B.15}$$

Substitution of these equations to the boundary conditions Eq.(B.2) or to (B.6) yields

$$\varepsilon_0\varepsilon_S E_S\left(2+\frac{2R}{R_d}+\frac{R^2}{R_d^2}\right)\cos\theta-\varepsilon_0\varepsilon_b E_i\cos\theta-P\cos\theta=0. \quad\Rightarrow \varepsilon_S\left(2+\frac{2R}{R_d}+\frac{R^2}{R_d^2}\right)E_S-\varepsilon_b E_i=\frac{P}{\varepsilon_0} \tag{B.16}$$

The solution of the linear system of Eqns.(B.13) and (B.16) have the form:

$$E_i=-\frac{P}{\varepsilon_0}\frac{1}{\varepsilon_b+2\varepsilon_S+\varepsilon_S R^2/\left(R_d^2+R_d R\right)}\approx-\frac{P}{\varepsilon_0}\frac{1}{\varepsilon_b+2\varepsilon_S+\varepsilon_S(R/R_d)}, \tag{B.17a}$$

$$E_S=\frac{P}{\varepsilon_0}\frac{R_d/(R_d+R)}{\varepsilon_b+2\varepsilon_S+\varepsilon_S R^2/\left(R_d^2+R_d R\right)}\approx\frac{P}{\varepsilon_0}\frac{R_d/R}{\varepsilon_b+2\varepsilon_S+\varepsilon_S(R/R_d)}. \tag{B.17b}$$

Approximate equalities in Eq.(B.17) are valid at $R_d<<R$.

Below we use the expression (B.17) for the formulation of the phenomenological equations of state. Substituting Eqs.(B.17) into the LGD equation one obtains the equation for the spontaneous polarization of the spherical particle

$$\left(\alpha_T\left(T-T_C^*\right)+\frac{1}{\varepsilon_0\left(\varepsilon_b+2\varepsilon_S+\varepsilon_S R^2/\left(R_d^2+R_d R\right)\right)}\right)P+\beta P^3=0 \tag{B.18}$$

Approximate expression for the nanoparticle transition temperature $T_{cr}$ from SDFE to PE phase follows from Eq.(B.18):

$$T_{cr}(R)=T_C^{*}-\frac{1}{\alpha_T\varepsilon_0\left[\varepsilon_b+2\varepsilon_S+\varepsilon_S R^2/\left(R_d^2+R_d R\right)\right]}, \tag{B.19}$$

### B3. Approximate solution for anisotropic dielectric shell in an external field

In particular case of a dielectric anisotropic shell, $\varepsilon_\perp \neq \varepsilon_{||}$ and $R_d \to \infty$ the solution of Eq.(B.3c) is

$$f(r)=E_S^-\left(\frac{r}{R}\right)^{\mu_1}+E_S^+\left(\frac{r}{R}\right)^{\mu_2}, \quad \mu_1=-\frac{1}{2}+\frac{1}{2}\sqrt{1+8\frac{\varepsilon_\perp}{\varepsilon_{||}}}, \quad \mu_2=-\frac{1}{2}-\frac{1}{2}\sqrt{1+8\frac{\varepsilon_\perp}{\varepsilon_{||}}}. \tag{B.20}$$

$$\varphi_i=-E_i r\cos\theta, \qquad 0\le r<R \tag{B.21a}$$

$$\varphi_S=-\left(E_S^-\left(\frac{r}{R}\right)^{\mu_1}+E_S^+\left(\frac{r}{R}\right)^{\mu_2}\right)R\cos\theta, \qquad R<r<R+\Delta R. \tag{B.21b}$$

$$\varphi_e=\left(E_e\frac{R^3}{r^2}-r\,E_{ext}\right)\cos\theta, \qquad r>R+\Delta R. \tag{B.21c}$$

Four constants $E_i$, $E_S^{\pm}$, and $E_e$ should be determined from the boundary conditions (B.2). Namely the potential continuity at $r=R$ and $r=R+\Delta R$ gives two equations:

$$E_i=E_S^+ + E_S^-, \tag{B.22a}$$

$$E_S^-\left(\frac{R+\Delta R}{R}\right)^{\mu_1}+E_S^+\left(\frac{R+\Delta R}{R}\right)^{\mu_2}=\left(\frac{R+\Delta R}{R}\right)E_{ext}-\frac{R^2E_e}{(R+\Delta R)^2}. \tag{B.22b}$$

Using the expressions for the radial components of electric displacements, $(\mathbf{D}_i)_r=(\varepsilon_0\varepsilon_b E_i+P)\cos\theta$, $(\mathbf{D}_S)_r=\varepsilon_0\varepsilon_{||}\left[E_S^-\mu_1\left(\frac{r}{R}\right)^{\mu_1-1}+E_S^+\mu_2\left(\frac{r}{R}\right)^{\mu_2-1}\right]\cos\theta$, and $(\mathbf{D}_e)_r=\varepsilon_0\varepsilon_e\left[2E_e\frac{R^3}{r^3}+E_{ext}\right]\cos\theta$, from the conditions on the normal components electric displacements continuity, we obtained two other equations:

$$\varepsilon_0\varepsilon_{||}\left[E_S^-\mu_1+E_S^+\mu_2\right]=\varepsilon_0\varepsilon_b E_i+P, \tag{B.22c}$$

$$\varepsilon_e\left[\frac{2E_eR^3}{(R+\Delta R)^3}+E_{ext}\right]=\varepsilon_{||}\left[E_S^-\mu_1\left(\frac{R+\Delta R}{R}\right)^{\mu_1-1}+E_S^+\mu_2\left(\frac{R+\Delta R}{R}\right)^{\mu_2-1}\right] \tag{B.22d}$$

It follows from Eq.(B.22a) that $E_i-E_S^+=E_S^-$ and $\varepsilon_0\varepsilon_{||}\left[(E_i-E_S^+)\mu_1+E_S^+\mu_2\right]=\varepsilon_0\varepsilon_b E_i+P$. Therefore, using $E_S^+=\frac{\varepsilon_b-\mu_1\varepsilon_{||}}{\varepsilon_{||}(\mu_2-\mu_1)}E_i+\frac{P}{\varepsilon_0\varepsilon_{||}(\mu_2-\mu_1)}$, and $E_S^-=-\frac{\varepsilon_b-\mu_2\varepsilon_{||}}{\varepsilon_{||}(\mu_2-\mu_1)}E_i-\frac{P}{\varepsilon_0\varepsilon_{||}(\mu_2-\mu_1)}$, and combining (B.22b) and (B.22d), one could find the constant $E_e$ from the equation

$\left(E_S^-\left(\frac{R+\Delta R}{R}\right)^{\mu_1-1}+E_S^+\left(\frac{R+\Delta R}{R}\right)^{\mu_2-1}\right)=E_{ext}-\frac{R^3E_e}{(R+\Delta R)^3}$ . Finally from Eq.(B.22d) one could get the following equation for $E_i$:

$$\begin{gathered}\left(\frac{\varepsilon_b-\mu_1\varepsilon_{||}}{\varepsilon_{||}(\mu_2-\mu_1)}E_i+\frac{P}{\varepsilon_0\varepsilon_{||}(\mu_2-\mu_1)}\right)\left(\frac{\varepsilon_{||}}{\varepsilon_e}\mu_2+2\right)\left(\frac{R+\Delta R}{R}\right)^{\mu_2-1}\\-\left(\frac{\varepsilon_b-\mu_2\varepsilon_{||}}{\varepsilon_{||}(\mu_2-\mu_1)}E_i+\frac{P}{\varepsilon_0\varepsilon_{||}(\mu_2-\mu_1)}\right)\left(\frac{\varepsilon_{||}}{\varepsilon_e}\mu_1+2\right)\left(\frac{R+\Delta R}{R}\right)^{\mu_1-1}=3E_{ext}\end{gathered}. \qquad \text{(B.23a)}$$

The equation gives the following expression for the internal electric field $E_i$:

$$E_i=\frac{3\varepsilon_{||}\varepsilon_e(\mu_2-\mu_1)E_{ext}+\left((\varepsilon_{||}\mu_1+2\varepsilon_e)\left(\frac{R+\Delta R}{R}\right)^{\mu_1-1}-(\varepsilon_{||}\mu_2+2\varepsilon_e)\left(\frac{R+\Delta R}{R}\right)^{\mu_2-1}\right)\frac{P}{\varepsilon_0}}{(\varepsilon_b-\mu_1\varepsilon_{||})(\varepsilon_{||}\mu_2+2\varepsilon_e)\left(\frac{R+\Delta R}{R}\right)^{\mu_2-1}-(\varepsilon_b-\mu_2\varepsilon_{||})(\varepsilon_{||}\mu_1+2\varepsilon_e)\left(\frac{R+\Delta R}{R}\right)^{\mu_1-1}} \qquad \text{(B.23b)}$$

Since

$$E_i^{ext}(R,\Delta R)=\frac{3\varepsilon_{||}\varepsilon_e(\mu_2-\mu_1)\left(\frac{R+\Delta R}{R}\right)^{1-\mu_1}}{1-\frac{\varepsilon_{||}\mu_2+2\varepsilon_e}{\varepsilon_{||}\mu_1+2\varepsilon_e}\left(\frac{R+\Delta R}{R}\right)^{\mu_2-\mu_1}}\eta(R,\Delta R)E_{ext} \qquad \text{(B.24a)}$$

$$E_i^{dep}=-\frac{P}{\varepsilon_0}\eta(R,\Delta R), \qquad \text{(B.24b)}$$

where the depolarization factor is introduced as:

$$\begin{aligned}\eta(R,\Delta R)&=\frac{1}{\varepsilon_b-\mu_1\varepsilon_{||}}\left[1-\frac{\varepsilon_{||}\mu_2+2\varepsilon_e}{\varepsilon_{||}\mu_1+2\varepsilon_e}\left(\frac{R+\Delta R}{R}\right)^{\mu_2-\mu_1}\right]\left[\frac{\varepsilon_{||}\mu_2+2\varepsilon_e}{\varepsilon_{||}\mu_1+2\varepsilon_e}\left(\frac{R+\Delta R}{R}\right)^{\mu_2-\mu_1}-\frac{\varepsilon_b-\mu_2\varepsilon_{||}}{\varepsilon_b-\mu_1\varepsilon_{||}}\right]^{-1}\\&\approx\begin{cases}\frac{1}{\varepsilon_b+2\varepsilon_e}, & \Delta R\to 0\\ \frac{1}{\varepsilon_b-\mu_2\varepsilon_{||}}\equiv\frac{2}{\varepsilon_{||}+\varepsilon_{||}\sqrt{1+8\frac{\varepsilon_\perp}{\varepsilon_{||}}}+2\varepsilon_b}, & \Delta R\to\infty\end{cases}\end{aligned} \qquad \text{(B.24c)}$$

One can introduce the "effective" permittivity from Eq.(B.24b) as $\varepsilon_{eff}^S=\frac{\varepsilon_{||}}{4}\left(1+\sqrt{1+8\frac{\varepsilon_\perp}{\varepsilon_{||}}}\right)$.

The expressions (B.24) can be used for the formulation of the phenomenological equations of state. Namely, substituting Eqs.(B.24) into the LGD equation one obtains the equation for the polarization of the spherical particle

$$\left(\alpha_T\left(T-T_C^*\right)+\frac{\eta(R,\Delta R)}{\varepsilon_0}\right)P+\beta P^3=E_i^{ext}(R,\Delta R) \qquad \text{(B.25)}$$

An approximate expression for the nanoparticle transition temperature $T_{cr}$ from SDFE to PE phase directly follows from Eq.(B.25):

$$T_{cr}(R)=T_C^{*}-\frac{\eta(R,\Delta R)}{\alpha_T\varepsilon_0}, \qquad \text{(B.26)}$$

**APPENDIX C. Derivation of PE-PDFE transition temperature for uniaxial ferroelectric nanoparticles**

The linearized system of equations for polarization and electric potential inside and outside the ferroelectric nanoparticle has the following form

$$\alpha P_3 - g_{11}\frac{\partial^2 P_3}{\partial z^2} - g_{44}\left(\frac{\partial^2 P_3}{\partial x^2}+\frac{\partial^2 P_3}{\partial y^2}\right) = -\frac{\partial \varphi}{\partial z}, \qquad \text{(C.1a)}$$

$$\left(\frac{\partial^2}{\partial x^2}+\frac{\partial^2}{\partial y^2}+\frac{\partial^2}{\partial z^2}\right)\varphi^{(in)}=\frac{1}{\varepsilon_0\varepsilon_b}\frac{\partial P_3}{\partial z}, \qquad \left(\frac{\partial^2}{\partial x^2}+\frac{\partial^2}{\partial y^2}+\frac{\partial^2}{\partial z^2}\right)\varphi^{(out)}=0, \qquad \text{(C.1b)}$$

with appropriate boundary conditions at the particle surface S:

$$\left.\left(\frac{\partial P_3}{\partial r}\right)\right|_{r=R}=0, \quad \left.\left(-\varepsilon_0\varepsilon_b\frac{\partial\varphi^{(in)}}{\partial r}+P_3+\varepsilon_0\varepsilon_e\frac{\partial\varphi^{(out)}}{\partial r}-\varepsilon_0\varepsilon_{eff}^{S}\frac{\varphi^{(in)}}{R_d}\right)\right|_{r=R}=0. \quad \left.\left(\varphi^{(out)}-\varphi^{(in)}\right)\right|_{r=R}=0 \qquad \text{(C.2)}$$

Hereinafter we suppose that the thickness of semiconducting shell satisfies the inequality $\Delta R \gg \frac{R_d}{\varepsilon_{eff}^{S}}$, where the "effective" dielectric permittivity $\varepsilon_{eff}^{S}$ is introduced in **Appendix B3.**

Let us consider harmonic-like fluctuations

$$P_3=P_k(z)\exp\left(i\vec{k}\,\vec{r}\right), \quad \varphi^{(in)}=\varphi_k^{(in)}(z)\exp\left(i\vec{k}\,\vec{r}\right), \quad \varphi^{(out)}=\varphi_k^{(out)}(z)\exp\left(i\vec{k}\,\vec{r}\right). \qquad \text{(C.3)}$$

Then the equations for amplitudes are as follows

$$\left(\alpha+g_{44}k^2\right)P_k - g_{11}\frac{\partial^2 P_k}{\partial z^2}=-\frac{\partial\varphi_k}{\partial z}, \qquad \text{(C.3a)}$$

$$\frac{\partial^2\varphi_k^{(in)}}{\partial z^2}-k^2\varphi_k^{(in)}=\frac{1}{\varepsilon_0\varepsilon_b}\frac{\partial P_k}{\partial z}, \qquad \frac{\partial^2\varphi_k^{(out)}}{\partial z^2}-k^2\varphi_k^{(out)}=0. \qquad \text{(C.3b)}$$

where $k^2=k_x^2+k_y^2$. Differentiation of the Eqs. (C.3) gives

$$\left(\frac{\partial^2}{\partial z^2}-k^2\right)\left[\left(\alpha+g_{44}k^2\right)P_k - g_{11}\frac{\partial^2 P_k}{\partial z^2}\right]=\left(\frac{\partial^2}{\partial z^2}-k^2\right)\left[-\frac{\partial\varphi_k}{\partial z}\right], \qquad \text{(C.4a)}$$

$$\frac{\partial}{\partial z}\left(\frac{\partial^2}{\partial z^2}-k^2\right)\varphi_k^{(in)}=\frac{1}{\varepsilon_0\varepsilon_b}\frac{\partial^2 P_k}{\partial z^2}. \qquad \text{(C.4b)}$$

Hence, one could exclude the potential amplitude from Eq.(C.4a) and get a single equation for polarization amplitude in the form:

$$\left(\frac{\partial^2}{\partial z^2}-k^2\right)\left[\left(\alpha+g_{44}k^2\right)P_k-g_{11}\frac{\partial^2 P_k}{\partial z^2}\right]=-\frac{1}{\varepsilon_0\varepsilon_b}\frac{\partial^2 P_k}{\partial z^2} \qquad \text{(C.5)}$$

Let us look for the solution of (C.5) in the form $P_3\sim\exp(q\,z)$, where inverse characteristic length $w$ satisfies the following equation:

$$q^4-\left(\frac{\alpha+g_{44}k^2}{g_{11}}+k^2+\frac{1}{\varepsilon_0\varepsilon_b g_{11}}\right)q^2+\frac{\alpha+g_{44}k^2}{g_{11}}k^2=0 \qquad \text{(C.6a)}$$

Its solutions could be written as

$$q_{1,2}^2=\frac{1}{2}\left(\frac{\alpha+g_{44}k^2}{g_{11}}+k^2+\frac{1}{\varepsilon_0\varepsilon_b g_{11}}\pm\sqrt{\left(\frac{\alpha+g_{44}k^2}{g_{11}}+k^2+\frac{1}{\varepsilon_0\varepsilon_b g_{11}}\right)^2-4\frac{\alpha+g_{44}k^2}{g_{11}}k^2}\right) \qquad \text{(C.6b)}$$

It should be noted, that in most cases $\varepsilon_0\varepsilon_b g_{11}<<\left\{1/k^2,\,g_{11}/|\alpha|,\,g_{44}/|\alpha|\right\}$, hence the following approximations are valid

$$q_1\approx\sqrt{\frac{\left(\alpha+g_{44}k^2\right)k^2}{\alpha+g_{44}k^2+g_{11}k^2+\dfrac{1}{\varepsilon_0\varepsilon_b}}},\quad q_2\approx\frac{1}{\sqrt{\varepsilon_0\varepsilon_b g_{11}}} \qquad \text{(C.6c)}$$

Now we could write the general solution of Eq.(C.5) in the form:

$$P_3=s_1\sinh(q_1z)+s_2\sinh(q_2z)+c_1\cosh(q_1z)+c_2\cosh(q_2z) \qquad \text{(C.7a)}$$

The four constants $s_i$ and $c_i$ should be found from the boundary conditions (C.2). Formal solution is zero, since we have a system of homogeneous linear equations for $s_i$ and $c_i$; but we are interested in the stability analysis, hence we should look for the zero point of the corresponding determinant of the linear equations system for $s_i$ and $c_i$. Since counter domain walls are charged and hence have much higher energy in comparison to parallel ones, the antisymmetric part of the solution corresponding to nonzero $s_i$ is always unstable from energetic considerations.

Unfortunately exact solution for the constants $c_i$ in the functions $P_3=c_1\cosh(q_1z)+c_2\cosh(q_2z)$ and constants $f_i$, $f$ in electric potentials $\varphi_k^{(in)}=f_1\sinh(q_1z)+f_2\sinh(q_2z)$ and $\varphi_k^{(out)}=f\exp(-|k|(z-h))$ are impossible to find in a finite form. Assuming that in the vicinity of the particle poles $z=\pm R$, the curvature of the spherical surface can be neglected, we obtained the system of four linear equations

$$c_1q_1\sinh(q_1h)+c_2q_2\sinh(q_2h)=0,\quad f_1\sinh(q_1h)+f_2\sinh(q_2h)-f=0,\quad c_i=\varepsilon_0\varepsilon_b\frac{q_i^2-k^2}{q_i}f_i \qquad \text{(C.8a)}$$

$$-\varepsilon_0\varepsilon_b\left(q_1f_1\cosh(q_1h)+q_2f_2\cosh(q_2h)\right)+\left(c_1\cosh(q_1h)+c_2\cosh(q_2h)\right)-|k|\varepsilon_0\varepsilon_e f-\varepsilon_0\frac{\varepsilon_{eff}^S}{R_d}f=0. \qquad \text{(C.8b)}$$

After cumbersome calculations, the conditions of zero determinant of the system (C.8) can be simplified under the validity of strong inequalities $|k|\varepsilon_e \frac{R_d}{\varepsilon_{eff}^S} << 1$. The latter strong inequality is valid for most cases, therefore, recalling the condition $q_2 >> |k|$,

$$\alpha + g_{44}k^2 + \frac{1}{\varepsilon_0}\left(\left(\varepsilon_b + 2\varepsilon_e\right) + \frac{\varepsilon_{eff}^S}{R_d}R + \left(\varepsilon_b + 2\varepsilon_e\right)\left(\xi k R\right)^2\right)^{-1} \approx 0. \qquad \text{(C.9a)}$$

Parameter $\xi$ is a geometrical factor that appeared to be close to ¼ for the considered core-and-shell model. This expression for the critical point should be further minimized with respect to the wave vector *k*:

$$\left(g_{44} - \frac{\left(\xi R\right)^2}{\varepsilon_0\left(\varepsilon_b + 2\varepsilon_e\right)}\left(1 + \frac{\varepsilon_{eff}^S R}{\left(\varepsilon_b + 2\varepsilon_e\right)R_d} + \left(\xi R \cdot k\right)^2\right)^{-2}\right)2k = 0. \qquad \text{(C.9a)}$$

Since zero root *k*=0 of Eq.(C.9) corresponds to a single domain state, we neglected it and obtained the following value of the domain structure wave vector at the transition point

$$k_{\min}\left(R\right) = \sqrt{\frac{1}{\sqrt{\varepsilon_0\left(\varepsilon_b + 2\varepsilon_e\right)g_{44}}\,\xi R} - \left(1 + \frac{\varepsilon_{eff}^S R}{\left(\varepsilon_b + 2\varepsilon_e\right)R_d}\right)\left(\frac{1}{\xi R}\right)^2} \qquad \text{(C.10)}$$

It is valid under the condition $\frac{\xi}{\sqrt{\varepsilon_0\left(\varepsilon_b + 2\varepsilon_e\right)g_{44}}} \geq \frac{1}{R} + \frac{\varepsilon_{eff}^S}{\left(\varepsilon_b + 2\varepsilon_e\right)R_d}$. With respect to Eq. (C.10), Eq.(C.8) could be further simplified to $\alpha + \frac{\sqrt{\varepsilon_0\varepsilon_b g_{44}}}{\varepsilon_0\varepsilon_b}\frac{2}{\xi R} - g_{44}\left(1 + \frac{\varepsilon_{eff}^S R}{\varepsilon_b R_d}\right)\left(\frac{1}{\xi R}\right)^2 \approx 0$.

Hence an approximate analytical expression for the transition temperature of the spherical nanoparticle from the PDFE to PE phase is:

$$T_{PE-PDFE}\left(R\right) \approx T_C^*\left(R\right) - \frac{1}{\alpha_T}\left(g_{44}k_{\min}^2 + \frac{\varepsilon_0^{-1}}{\left(\varepsilon_{eff}^S R/R_d\right) + \left(\varepsilon_b + 2\varepsilon_e\right)\left(1 + k_{\min}^2\left(\xi R\right)^2\right)}\right). \qquad \text{(C.11a)}$$

Here the first term originated from the correlation effect and the second one is from depolarization field energy of the domain stripes. Its origin is related to the corresponding spherical eigenfunctions. Corresponding wave vector $k_{\min}$ of the domain structure onset are

$$k_{\min} = \sqrt{\frac{1}{\sqrt{\varepsilon_0\left(\varepsilon_b + 2\varepsilon_e\right)g_{44}}\,\xi R} - \left(1 + \frac{\varepsilon_{eff}^S R}{\left(\varepsilon_b + 2\varepsilon_e\right)R_d}\right)\left(\frac{1}{\xi R}\right)^2}, \qquad \text{(C.11b)}$$

Expressions (C.11) have physical sense under the condition

$$\frac{\xi}{\sqrt{\varepsilon_0\left(\varepsilon_b + 2\varepsilon_e\right)g_{44}}} \geq \frac{1}{R} + \frac{\varepsilon_{eff}^S}{\left(\varepsilon_b + 2\varepsilon_e\right)R_d} \qquad \text{(C.12)}$$

The fulfillment of the equality

$$\frac{1}{R}+\frac{\varepsilon_{eff}^{S}}{\left(\varepsilon_b+2\varepsilon_e\right)R_d}=\frac{\xi}{\sqrt{\varepsilon_0\left(\varepsilon_b+2\varepsilon_e\right)g_{44}}}\,. \tag{C.13}$$

corresponds to the tricritical point on the phase diagram in coordinates e.g. $T$ and $R_d$.